\documentclass[twocolumn]{aastex63}
\usepackage{color,comment}
\usepackage[normalem]{ulem}
\usepackage[caption=false]{subfig}
\usepackage{graphicx}
\usepackage{verbatim}
\usepackage{mathtools} 
\usepackage{float}
\usepackage{wrapfig}
\usepackage[splitrule,bottom]{footmisc}

\newcommand{\GG}[1]{} 

\received{}
\revised{}
\accepted{}

\shorttitle{How to dust temperature?}
\shortauthors{S. Lower et al.}

\begin{document}

\title{Cosmic Sands II: Challenges in Predicting and Measuring High-z Dust Temperatures}

\author[0000-0003-4422-8595]{Sidney Lower}
\affil{Department of Astronomy, University of Florida, 211 Bryant Space Science Center, Gainesville, FL, 32611, USA}
\author[0000-0002-7064-4309]{Desika Narayanan}
\affil{Department of Astronomy, University of Florida, 211 Bryant Space Science Center, Gainesville, FL, 32611, USA}
\affil{University of Florida Informatics Institute, 432 Newell Drive, CISE Bldg E251 Gainesville, FL, 32611, US}
\affil{Cosmic Dawn Centre at the Niels Bohr Institue, University of Copenhagen and DTU-Space, Technical University of Denmark}
\author[0000-0002-9235-3529]{Chia-Yu Hu}
\affil{Department of Astronomy, University of Florida, 211 Bryant Space Science Center, Gainesville, FL, 32611, USA}
\author[0000-0003-3474-1125]{George C. Privon}
\affil{National Radio Astronomy Observatory, 520 Edgemont Road, Charlottesville, VA 22903, USA}
\affil{Department of Astronomy, University of Florida, 211 Bryant Space Science Center, Gainesville, FL, 32611, USA}
\affil{Department of Astronomy, University of Virginia,
530 McCormick Road, Charlottesville, VA 22904, USA}

\begin{abstract}
In the current era of high-z galaxy discovery with JWST and ALMA, our ability to study the stellar populations and ISM conditions in a diverse range of galaxies at Cosmic Dawn has rapidly improved. At the same time, the need to understand the current limitations in modeling galaxy formation processes and physical properties in order to interpret these observations is critical. Here, we study the challenges in modeling galaxy dust temperatures, both in the context of forward modeling galaxy spectral properties from a hydrodynamical simulation and via backwards modeling galaxy physical properties from mock observations of far-infrared dust emission. Using the {\sc simba} model for galaxy formation combined with {\sc powderday} radiative transfer, we can accurately predict the evolution of dust at high redshift, though several aspects of the model are essentially free parameters (dust composition, sub-resolution dust in star-forming regions) that dull the predictive power of the model dust temperature distributions. We also highlight the uncertainties in the backwards modeling methods, where we find the commonly used models and assumptions to fit FIR SEDs and infer dust temperatures (e.g., single temperature, optically thin modified blackbody) largely fail to capture the complexity of high-z dusty galaxies. We caution that conclusions inferred from both simulations -- limited by resolution and post-processing techniques -- and observations -- limited by sparse data and simplistic model parameterizations -- are susceptible to unique and nuanced uncertainties that can limit the usefulness of current high-z dust measurements.

\end{abstract}

\section{Introduction}

 Like those of the gaseous interstellar medium (ISM), dust temperatures vary spatially within a galaxy, dependent on the proximity to star-forming regions, active galactic nuclei, and the galactic disk. Within a single galaxy there exists a dust temperature distribution, with the mass-weighted average temperature typically lower than the luminosity-weighted as the cold dust component dominates the mass while the warm dust dominates the luminosity. Characterizing the global `average' dust temperature is nontrivial; if a single dust temperature is used to describe a galaxy, the mass-weighted temperature is perhaps the most physical if concerned about characterizing the dust mass. On the other hand, the luminosity-weighted dust temperature is important for characterizing the intensity of star formation or AGN activity.

 In the context of galaxy observations, the global dust temperature is most commonly measured from galaxy spectral energy distributions (SEDs) and is significantly dependent on the assumptions made regarding the shape of the FIR SED \citep{yang_2007}. The most common model assumption is that the global dust emission can be well described by a single temperature modified blackbody; though in theory a multi-temperature model can more accurately describe the dust temperature distribution in galaxies and yield a better fit to the FIR SED, especially for IR luminous galaxies \citep{clements_2010}, the range and quality of photometry necessary to constrain such a model is challenging even for low redshift galaxies \citep{casey_2012, utomo_2019} in addition to the inherent degeneracies between the multi-temperature, optically thick, and variable spectral index solutions \citep{dunne_2001, klaas_2001, rangwala_2011}.
 
 The relationship between the SED-derived dust temperature and the real dust temperature distribution depends on both the model assumptions and the underlying dust properties (opacity, geometry, etc.) \citep{liang_lichen_2019}. In some cases, e.g., the standard assumption of optically thin, single temperature sources, correlations between model parameters can be induced, adding to the aforementioned uncertainties and degeneracies. These model dependencies and degeneracies between model parameters (e.g., dust temperature, opacity, and spectral index) are well established \citep{blain_2003, shetty_2009a, shetty_2009b, kelly_2012, juvela_2013} and can in principle be constrained with robust data spanning the range of FIR dust emission.

 Yet there is a scarcity of FIR constraints at high redshift: dust continuum measurements are difficult to obtain except for the rarest, brightest galaxies, and are typically tied to observations of FIR emission lines like [CII] or [O III] \citep[e.g.][]{pavesi_2016, bakx_2020, bakx_2021, harikane_2020, sommovigo_2022_rebels_alpine_dust, yoon_2022} and will inevitably compound the issues described above. As cautioned in \cite{spilker_2016} and \cite{drew_2022}, the lack of constraints on the dust column density make constraining the FIR optical depth properties of these galaxies nearly impossible. This can lead to considerable biases in the dust and gas masses measured from broadband SEDs \citep{cochrane_2022} as well as the potential evolution in dust temperatures across time \citep{drew_2022, sommovigo_2022_rebels_alpine_dust}. Moreover, the applicability of certain assumptions that may be valid for some low-z galaxies (e.g., optically thin FIR emission) is relatively unknown for high-z galaxies; though these assumptions are typical in the literature, their uncritical use can clearly lead to further uncertainties.

 In this paper, we aim to understand these uncertainties with the use of cosmological simulations. We employ the Cosmic Sands suite of early, dusty galaxies \citep{lower_2022_cosmic_sands} to probe the efficacy of existing methods to infer dust properties from observational methods. The Cosmic Sands simulations offer the advantage of a self-consistent dust evolution model, which we couple with {\sc powderday} dust radiative transfer \citep{powderday}, allowing us to forward model dust temperatures from the particle data and to backwards model dust temperatures as an observer would from the mock SEDs. Our goal is not to provide predictions for dust temperatures for high-z galaxies, but instead to use the Cosmic Sands galaxies as a testing ground for the methodologies used to infer dust properties in the literature.

This paper is organized as follows: in Section \S \ref{sec:numerical_methods}, we describe the Cosmic Sands suite of simulations based on the {\sc simba} galaxy formation model. In Section \S \ref{sec:mw_dust_temps}, we present an analysis of the radiative transfer derived dust temperatures of the Cosmic Sands galaxies and discuss the impact that the {\sc simba} dust model has on these results. In Section \S \ref{sec:measuring_dust_temps}, we describe and compare the techniques for measuring dust temperatures with a modified blackbody model and the relationship the SED-derived temperature measurement has with the physical dust temperatures. Finally, in Section \S \ref{sec:discussion}, we discuss the implications of our results for surveys of high-z dusty galaxies as well as the caveats to our galaxy formation model and post-processing radiative transfer. 

\section{Numerical Methods}\label{sec:numerical_methods}

In \cite{lower_2022_cosmic_sands}, we presented the Cosmic Sands suite of zoom-in cosmological simulations to study the build up of massive, dusty galaxies at Cosmic Dawn, focusing on the processes that impact efficient stellar mass build up. Here, we use the Cosmic Sands galaxies to study the various methods of measuring dust temperatures and what drives the diversity in galaxy averaged mass-weighted dust temperatures. Below, we briefly describe the {\sc simba} galaxy formation physics that Cosmic Sands is based on and the radiative transfer post-processing done on the simulation snapshots to calculate dust temperatures. 

\subsection{Cosmic Sands \& The {\sc simba} Galaxy Formation Model}\label{sec:simba}

Our galaxy sample is generated from the {\sc simba} galaxy formation model. {\sc Simba} \citep{dave_simba} is based on the {\sc Gizmo} gravity and hydrodynamics code \citep{hopkins_2015_gizmo} and includes models describing heating and cooling, star formation, chemical enrichment, feedback from stellar winds, dust production and growth, and blackhole (BH) accretion and feedback. We briefly summarize the key aspects of each model. 

\looseness=-1
Physically, star formation occurs in dense molecular clouds; numerically, the rate of star formation is dictated by the density of H$_2$ divided by the local dynamical timescale, with gas particles below a density of $n_\mathrm{H} = 0.13 \: \mathrm{cm}^{-3}$ blocked from star formation. The H$_2$ fraction is modeled with the sub-resolution prescription of \citet{krumholz_2011_h2} depending on the gas-phase metallicity and local gas column density.

\looseness=-1
{\sc Simba} uses the {\sc grackle-3} library \citep{smith_2017_grackle} to model radiative cooling and photoionization heating including a model for self-shielding based on \citet{rahmati_2013_self_shield}. The chemical enrichment model tracks $11$ elements from from Type Ia and II supernovae (SNe) and asymptotic giant branch (AGB) stars with yields following \cite{nomoto_simba_sne_yields}, \cite{iwamoto_1999_sne_yields}, and \cite{Oppenheimer_2006_agb_yields}, respectively.

\looseness=-1

The sub-resolution model for stellar feedback includes contributions from Type II SNe, radiation pressure, and stellar winds. The two-component stellar winds adopt the mass-loading factor scaling from {\sc fire} \citep{angles-alcazar_FIRE} with wind velocities given by \cite{muratov_2015}. Metal-enriched winds extract metals from nearby particles to represent the local enrichment by the SNe driving the wind. The sub-resolution model for feedback via active galactic nuclei (AGN) is implemented as a two-phase jet (low accretion rate) and radiative (high accretion rate) model. Thermal energy is injected into the surrounding interstellar medium (ISM) at high accretion rates, while BH-driven winds are produced at low accretion rates.

\looseness=-1
The dust model in {\sc simba} \citep{li_2019_dust} includes prescriptions for the production of dust by stellar sources, both in the remnants of Type II SNe explosions and in the metal-loaded winds of evolved AGB stars.  Dust can grow via metal accretion in the ISM, and be destroyed via thermal sputtering, supernova shocks, and astration in star-forming regions.   The dust is assumed to be at a fixed grain size of  $a=0.1$ $\mu$m, though we refer the reader to \cite{li_2020_mw_ext_curve} for an evolving grain size distribution model implemented into the {\sc simba} galaxy formation framework.

\begin{figure*}
    \includegraphics[width=\textwidth]{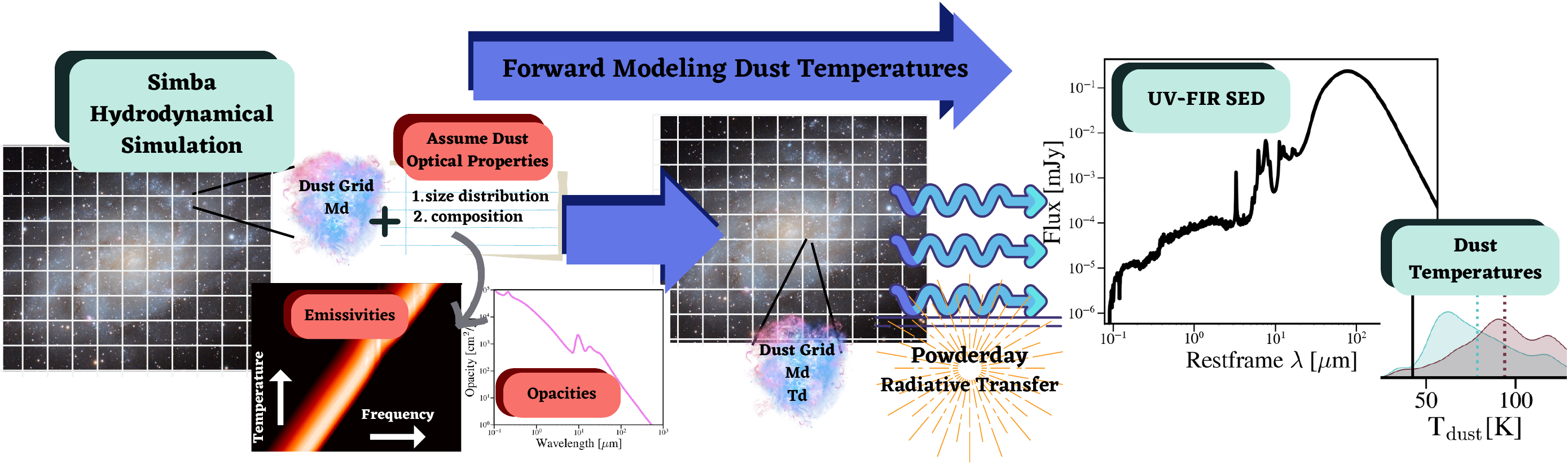}
    \caption{Schematic detailing the forward dust temperature modeling with {\sc powderday} radiative transfer. Starting from Cosmic Sands galaxies evolved with the {\sc simba} galaxy formation framework, we generate an octree grid onto which the gas particles (containing the dust mass information) are smoothed. We then assume dust grain properties (grain size distribution and composition) corresponding to the \cite{weingartner_draine_2001} $R_{\rm V} = 3.1$ extinction curve. This sets the emissivities and opacities of the dust in each cell. From there, the radiative transfer propagates through the dusty ISM, producing UV-FIR SEDs and dust temperature distributions for each galaxy.}
    \label{fig:pd_diagram}
\end{figure*}

The galaxies we use for this study are from the Cosmic Sands suite of cosmological zoom-in simulations presented in \cite{lower_2022_cosmic_sands}. These simulations were selected from dark matter only runs and re-run with baryons at higher particle counts within the selected halos, thereby creating a high-resolution ``zoom-in" of the halo of interest. The Cosmic Sands simulations were preferentially selected as the most massive halos in 32 distinct boxes with volumes of $25$ Mpc each. The baryonic mass resolution within the high-resolution region is $2.8\times10^5$ M$_{\odot}$. At redshift $z=6$, the halos span a range of masses (gas + stellar + dark matter) from $6\times10^{11}$ to $8\times10^{12}$ M$_{\odot}$. We use {\sc caesar} \citep{caesar} to identify the initial dark matter halos and later the galaxies associated with each halo with a 6D (position-velocity phase space) friends of friends finder (6D FOF). The spatial linking length for gas and star particles is 0.0056 times the mean interparticle spacing while the velocity linking length is 1 times the local velocity dispersion; gas and star particles that are bound in this way are associated with galaxies, from which the galaxy properties are calculated, which in turn are associated with their parent halos.

\begin{figure*}
    \centering
    \includegraphics[width=0.98\textwidth]{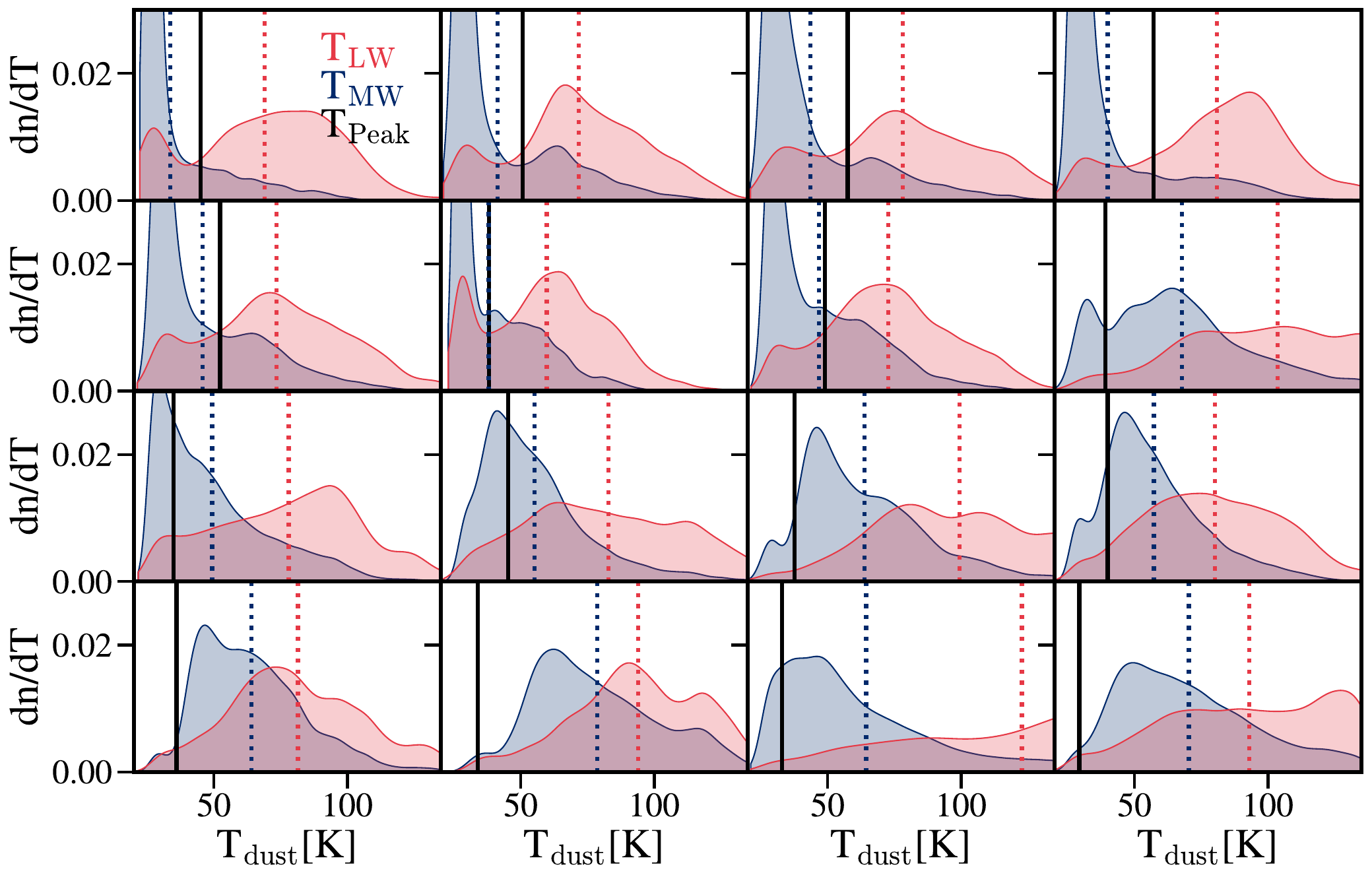}
    \caption{Dust temperature distributions for a selected sample of the Cosmic Sands galaxies, generated via {\sc powderday} radiative transfer. Galaxies increase in dust mass from the top-left to bottom-right subplots. The mass-weighted (luminosity-weighted) dust temperature distributions are shown in blue (red). The whole-galaxy mass-weighted and luminosity weighted temperatures are shown by the blue and red dotted lines. The black solid lines show the temperatures corresponding to the peak of the FIR SED.}
    \label{fig:temp_dist}
\end{figure*}

\subsection{Dust Radiative Transfer}\label{sec:rt}

\looseness=-1
We post-process the simulated galaxies with 3D dust radiative transfer (RT) to generate UV-FIR SEDs for each massive galaxy. Following the schematic in Figure \ref{fig:pd_diagram}, we use the radiative transfer code {\sc powderday} \citep{powderday}, which is built on {\sc YT} \citep{turk_2011_yt}, {\sc hyperion} \citep{robitaille_2011_hyperion}, and {\sc FSPS} \citep{fsps_1, fsps_2}, to construct the synthetic SEDs. For each galaxy, an octree is constructed and gas particles (containing the {\sc simba} dust mass information) are smoothed onto the grid which is refined to a maximum of 16 particles in each cell, with a minimum cell size ranging between $0.05 - 0.2$~kpc. With {\sc fsps}, we generate the intrinsic SEDs for the star particles within each cell using the stellar ages and metallicities as returned from the cosmological simulations. For these, we assume a \cite{kroupa_initial_2002} stellar IMF and the {\sc mist} stellar isochrones \citep{mist_1, mist_2}. These {\sc fsps} stellar SEDs are then propagated through the dusty ISM. 

 Polycyclic aromatic hydrocarbon (PAH) emission is included following the \citet{robitaille_pahs} model in which PAHs are assumed to occupy a constant fraction of the dust mass (here, modeled as grains with size $a<20$ \AA) and occupying $5.86\%$ of the dust mass. The diffuse dust distribution is derived from the on-the-fly self-consistent model of \cite{li_2019_dust}. As {\sc simba} contains a `passive' dust model, with a single grain size ($0.1\mu$m), we assume the extinction and emission models of \cite{draine_infrared_2007}. This model includes the \cite{weingartner_draine_2001} grain size distribution and the \cite{draine_03_araa} renormalization relative to hydrogen. Specifically, we assume the temperature and frequency dependent opacities and emissivities of the $R_{\rm V} \equiv A_{\rm V}/E(B-V) = 3.1$ curve from \cite{weingartner_draine_2001}, though the emissivities are parameterized in terms of the mean intensity absorbed by grains, rather than the average interstellar radiation field as in the original Draine \& Li model.

\looseness=-1
The radiative transfer propagates through the dusty ISM, with a resolution of $10^6$ photon packets, in a Monte Carlo fashion using {\sc hyperion}, which follows the \cite{lucy_rt} algorithm in order to determine the equilibrium dust temperature in each cell. We iterate until the energy absorbed by $99\%$ of the cells has changed by less than $1\%$. We generate SEDs corresponding to 25 random viewing angles for each galaxy, allowing us to probe the relationship between apparent dust temperatures and the intrinsic dust properties of each galaxy.

We note here an important detail relevant to forward modeling of dust temperatures in the following section: the emissivities described above, which account for the imperfect blackbody (greybody) nature of dust grain emission in the IR, are calculated for the specific grain sizes and compositions in the adopted model. While the emissivities are known for each dust cell, there is not a straightforward way to predict the emissivity of the \textit{composite} FIR SED as the emission also depends on the local temperature. Each grid cell containing dust will emit a blackbody with nonzero emissivity at the local temperature; the galaxy FIR SED is effectively the sum of these grey-bodies with an emissivity that will depend on the temperature distribution and the dust column density along the line-of-sight.

\section{Forward Modeling Dust Temperature}\label{sec:mw_dust_temps}

We derive the mass-weighted (T$_{\rm mass-weighted}$) and luminosity-weighted (T$_{\rm lum-weighted}$) dust temperatures for the Cosmic Sands galaxies using {\sc powderday} radiative transfer, with the full temperature distributions shown in Figure \ref{fig:temp_dist}. Because Cosmic Sands uses the {\sc simba} galaxy formation model that includes dust evolution, we do not have to make assumptions about the distribution of dust or its scaling with the gas or metal mass. The mass-weighted temperatures are calculated in each {\sc powderday} grid cell $i$ occupied by dust. We calculate T$_{\rm mass-weighted}$ as

\begin{equation}
    {\rm T}_{\rm mass-weighted} = \frac{\sum {\rm T}_{i} {\rm M}_{i}}{\sum {\rm M}_i},
\label{eq:T_MW}
\end{equation}
where T$_i$ and M$_i$ are the dust temperature and mass in each cell, respectively.

To compute the luminosity-weighted temperatures we instead weight by the luminosity in each cell $i$, assuming the dust emits as a modified blackbody at the temperature in the cell with absorption and emissivity values from \cite{weingartner_draine_2001}:

\begin{equation}
    \begin{split}
    &{\rm L}_i = \int_{8\mu m}^{1000\mu m} (1 - e^{-\tau_i}) (B_{\lambda} ({\rm T}_i) - B_{\lambda} ({\rm T_{CMB}})) d\lambda\\
    &{\rm T}_{\rm lum-weighted} = \frac{\sum {\rm T}_{i} {\rm L}_{i}}{\sum {\rm L}_i},
    \end{split}
\label{eq:T_LW}
\end{equation}
where $B_{\lambda}$ is the blackbody function. We note here that neither the mass-weighted nor luminosity-weighted temperatures are necessarily measurable quantities, as radiative transfer effects can bias temperatures measured from dust luminosities. To demonstrate this, we also plot in Figure \ref{fig:temp_dist} the apparent dust temperatures corresponding to the peak of the FIR SED (T$_{\rm peak}$) in the solid black lines (calculated from the spectral flux density $F_{\nu}$). For galaxies with lower dust mass, the peak temperature is typically similar to the luminosity-weighted temperature but the relationship between the peak temperature and the temperature distributions of dustier galaxies is more complex. As we will discuss in subsequent sections, namely \S\ref{sec:optical_depth}, this is primarily driven by optical depth effects in the FIR becoming more significant in dustier galaxies, including complex star-dust geometries with highly opaque dust column densities.

\begin{figure*}
    \centering
    \includegraphics[width=0.98\textwidth]{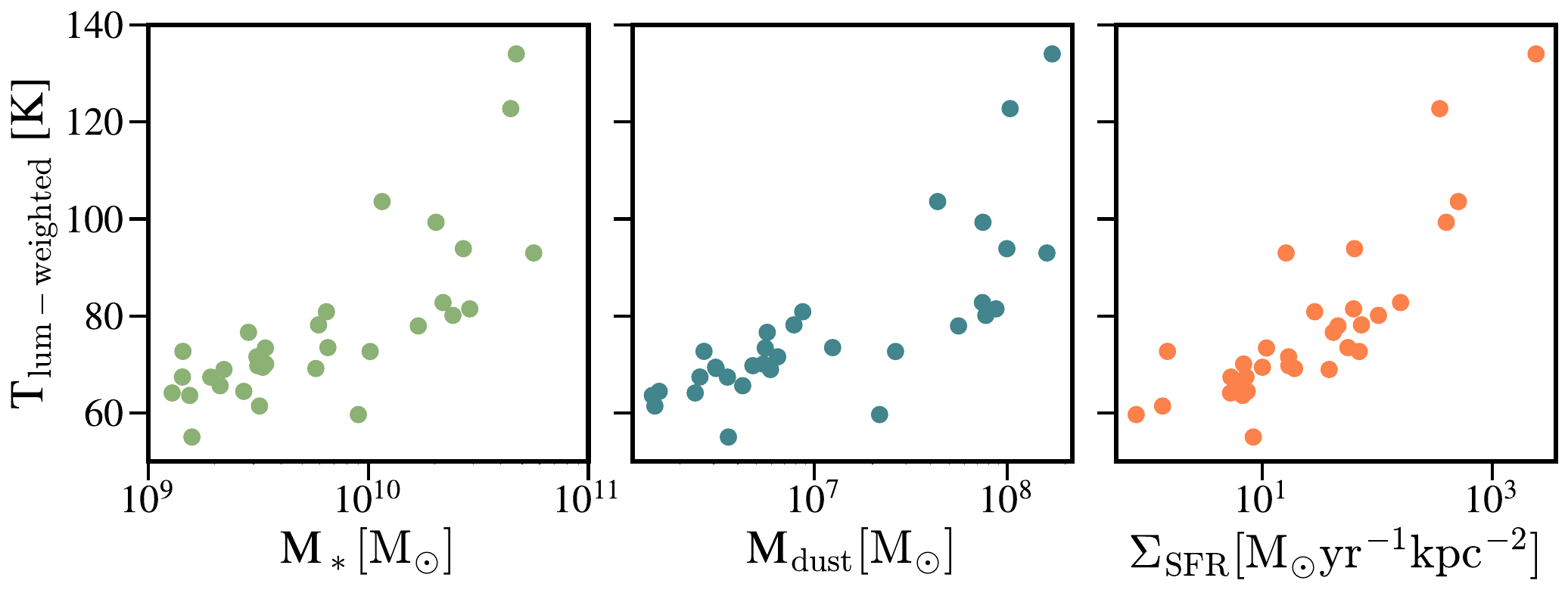}
    \caption{Luminosity-weighted dust temperatures as a function of total stellar mass, dust mass, and SFR surface density at $z=6.5$.}
    \label{fig:lum_weighted_temp}
\end{figure*}

\begin{figure*}
    \centering
    \includegraphics[width=0.95\textwidth]{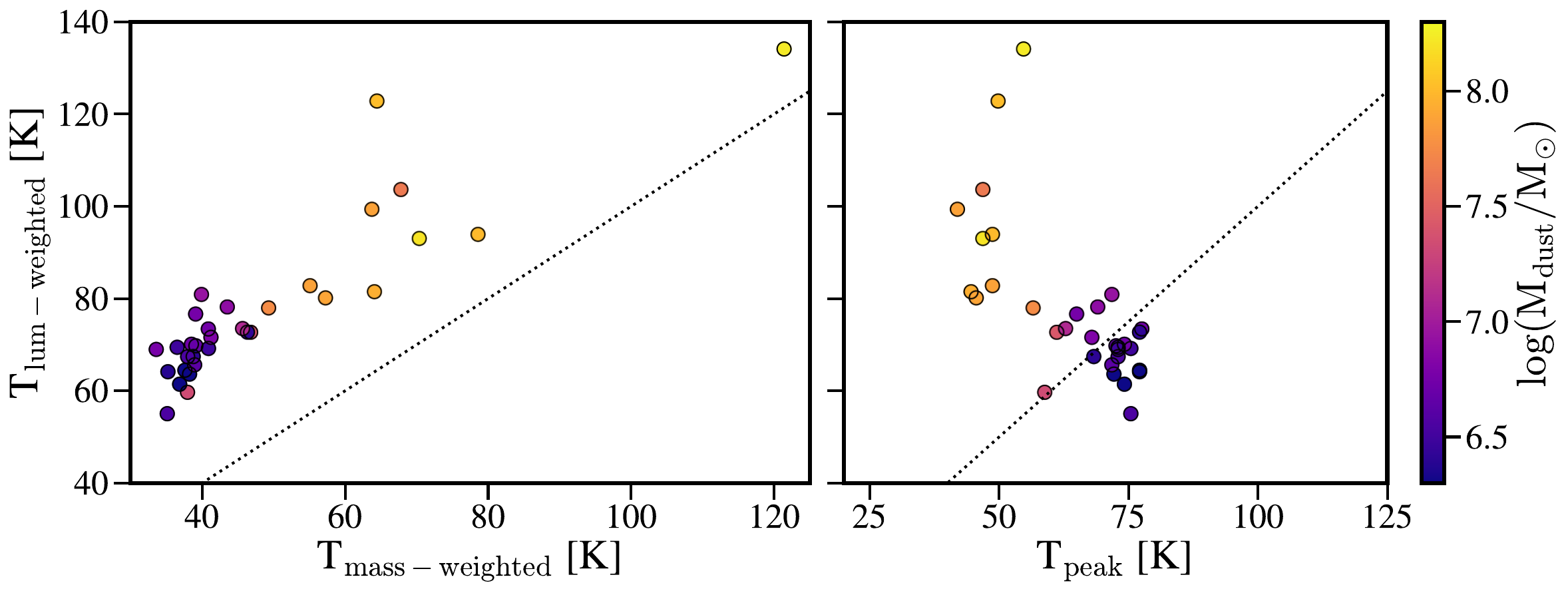}
    \caption{Luminosity-weighted dust temperatures as a function of mass-weighted temperature (left) and temperature corresponding to the peak of the SED (right). Points are color-coded by galaxy dust mass.}
    \label{fig:Tmw_Tlw}
\end{figure*}

\begin{figure*}[t!]
    \centering
    \includegraphics[width=0.48\textwidth]{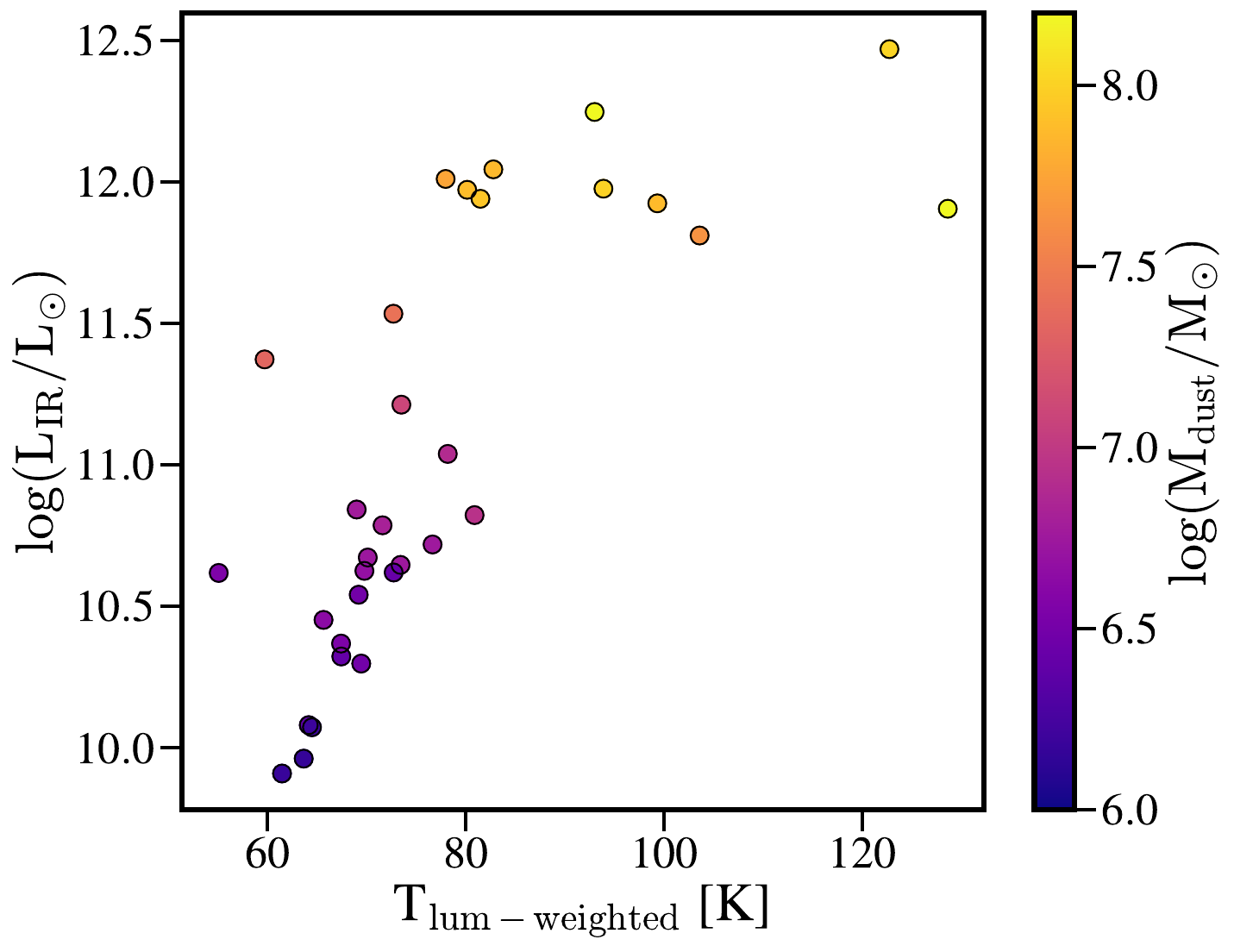}
    \includegraphics[width=0.48\textwidth]{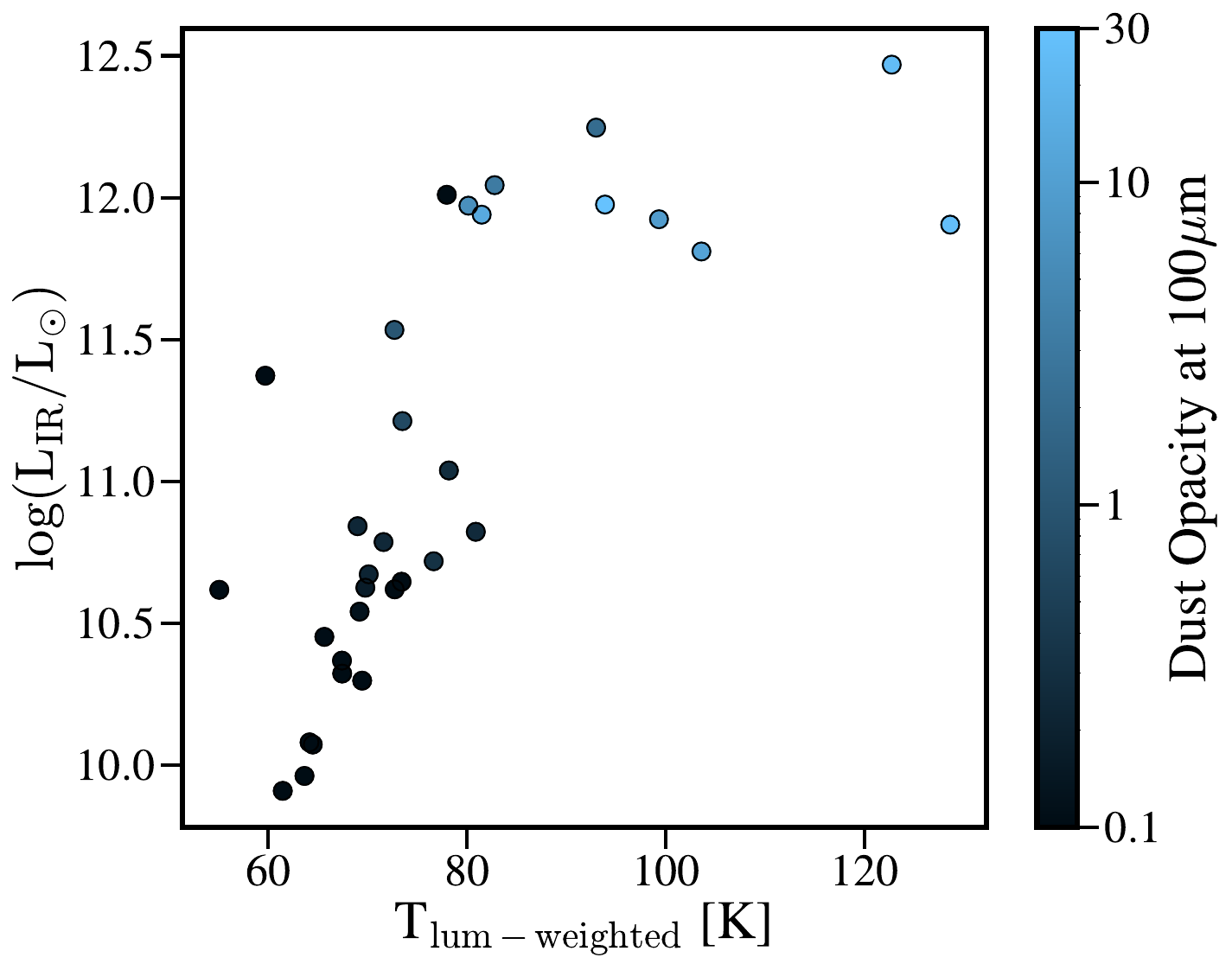}
    \caption{Infrared luminosity, calculated from {\sc powderday} SEDs chosen at a random viewing angle for each galaxy, as a function of luminosity-weighted dust temperatures. Points are color-coded by galaxy dust mass in the left panel and by dust optical depth at 100 $\mu$m in the right panel.}
    \label{fig:temp_lir_tau}
\end{figure*}

\begin{figure*}[t!]
    \centering
    \includegraphics[width=0.48\textwidth]{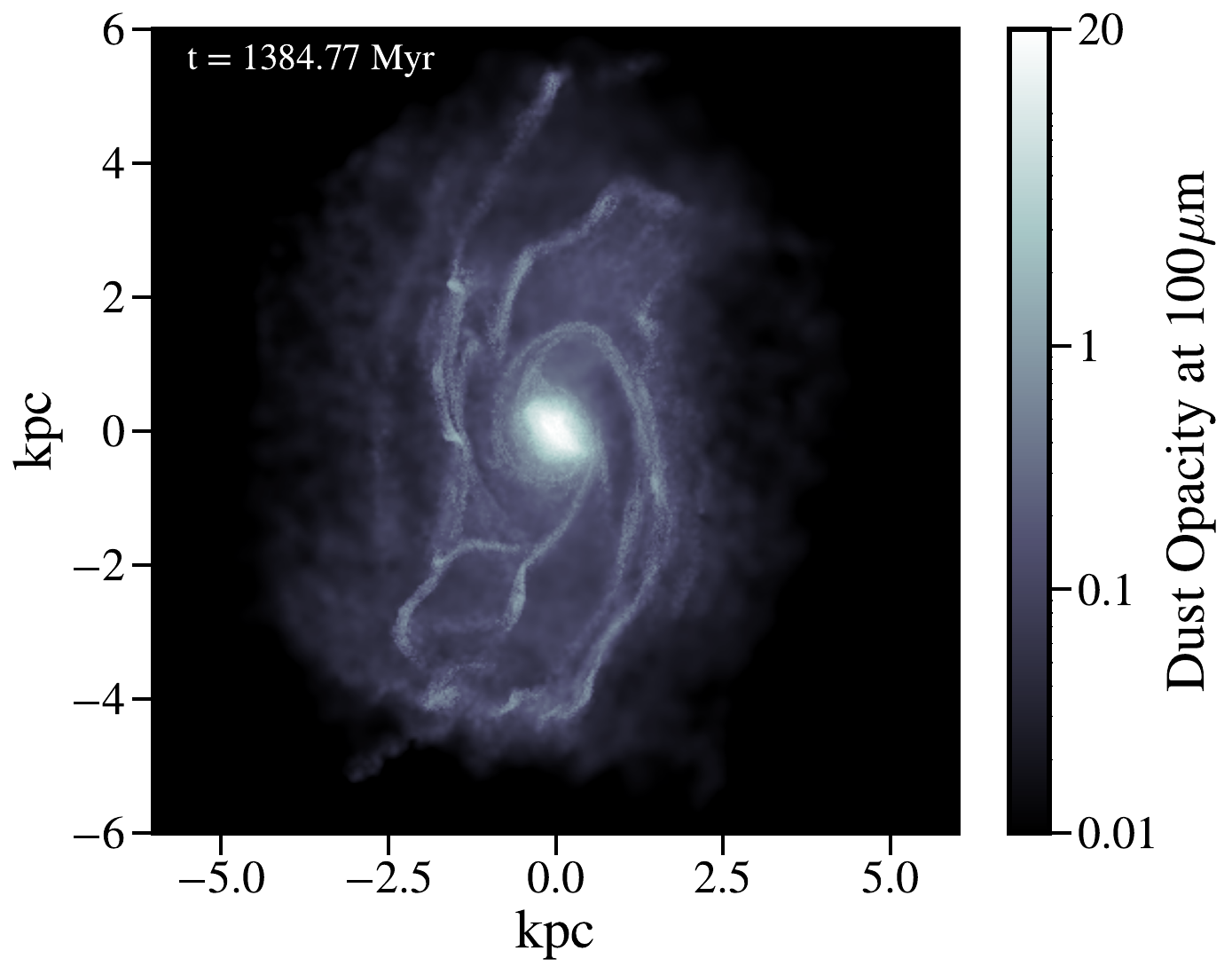}
    \includegraphics[width=0.48\textwidth]{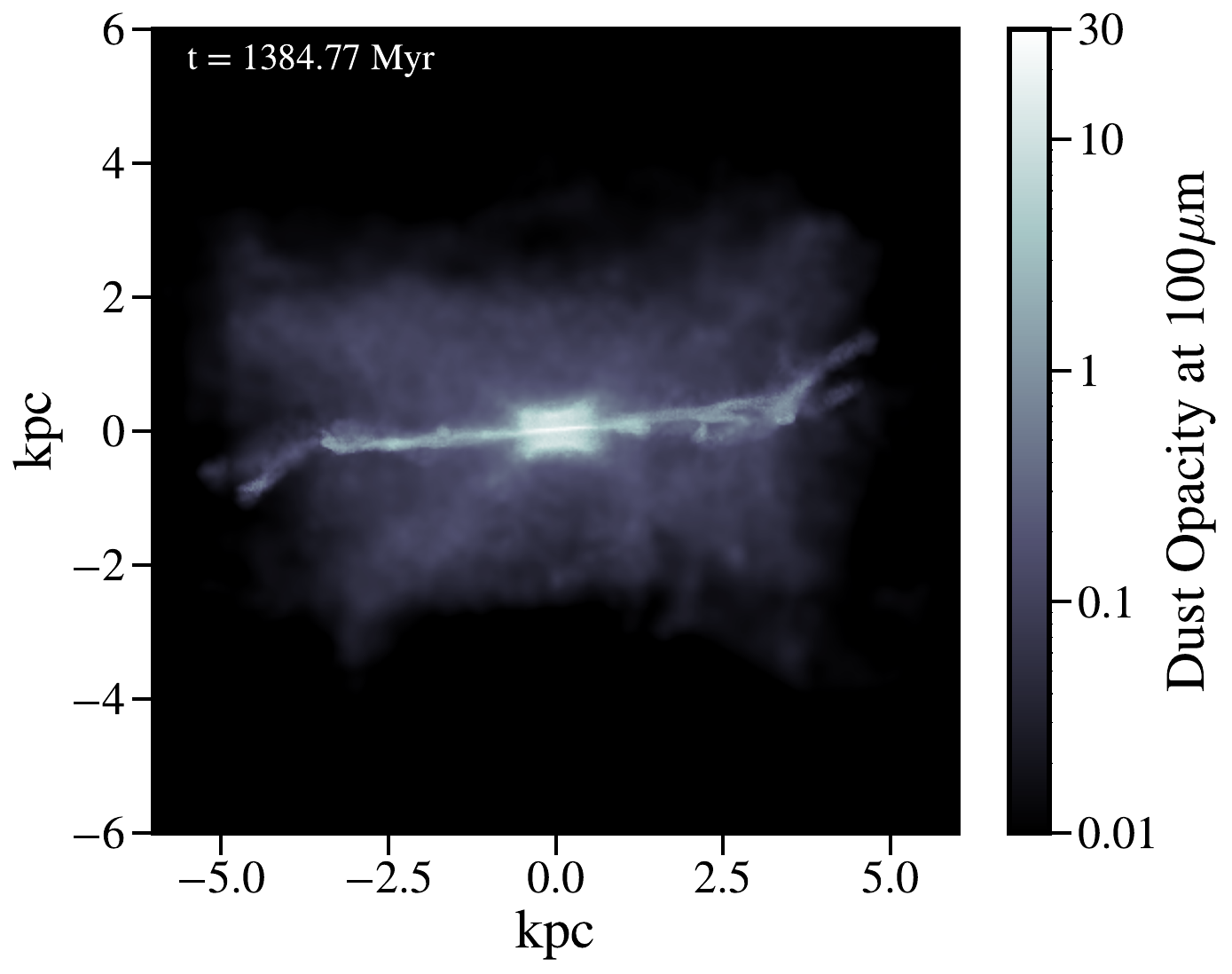}\\
    \includegraphics[width=0.48\textwidth]{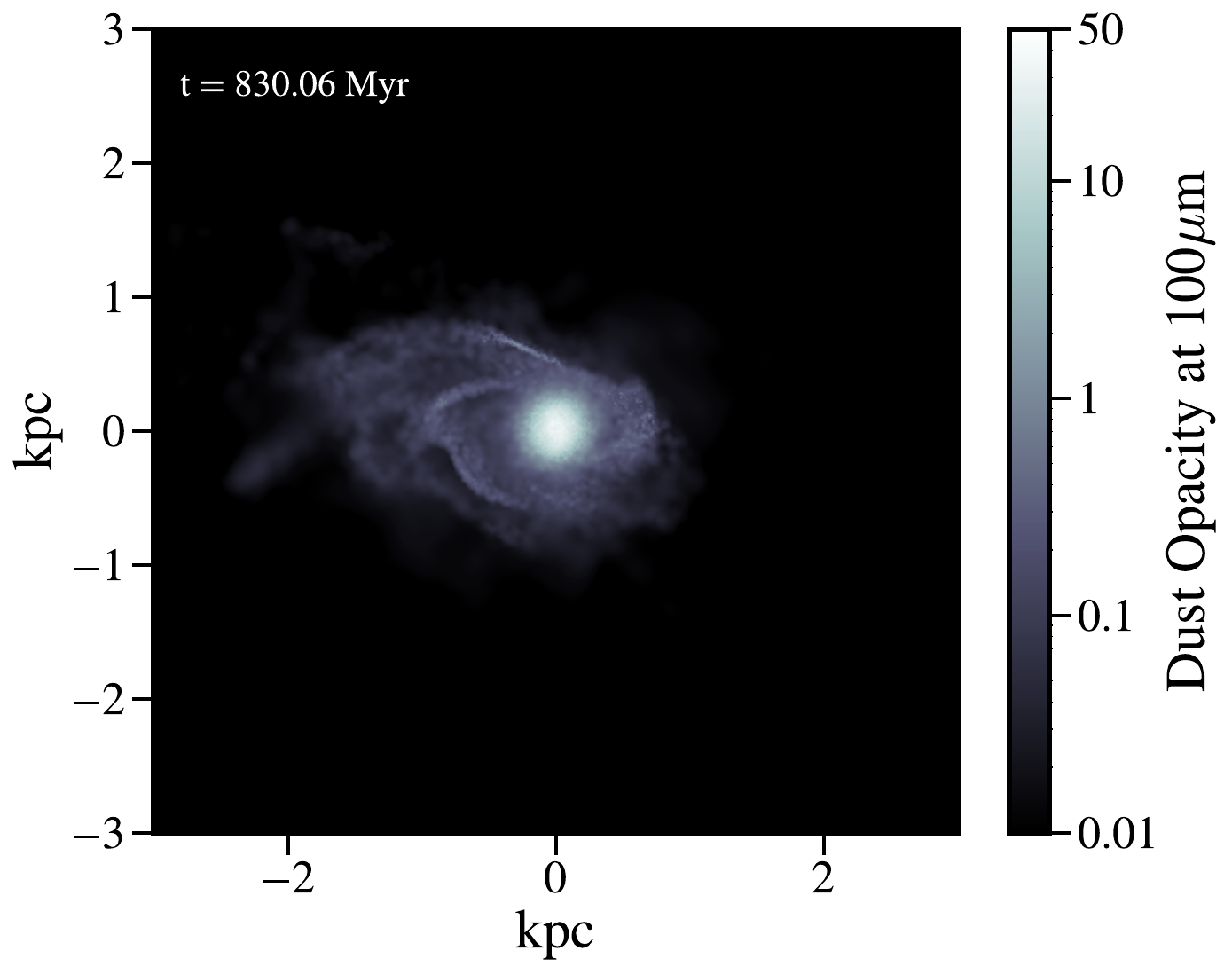}
    \includegraphics[width=0.48\textwidth]{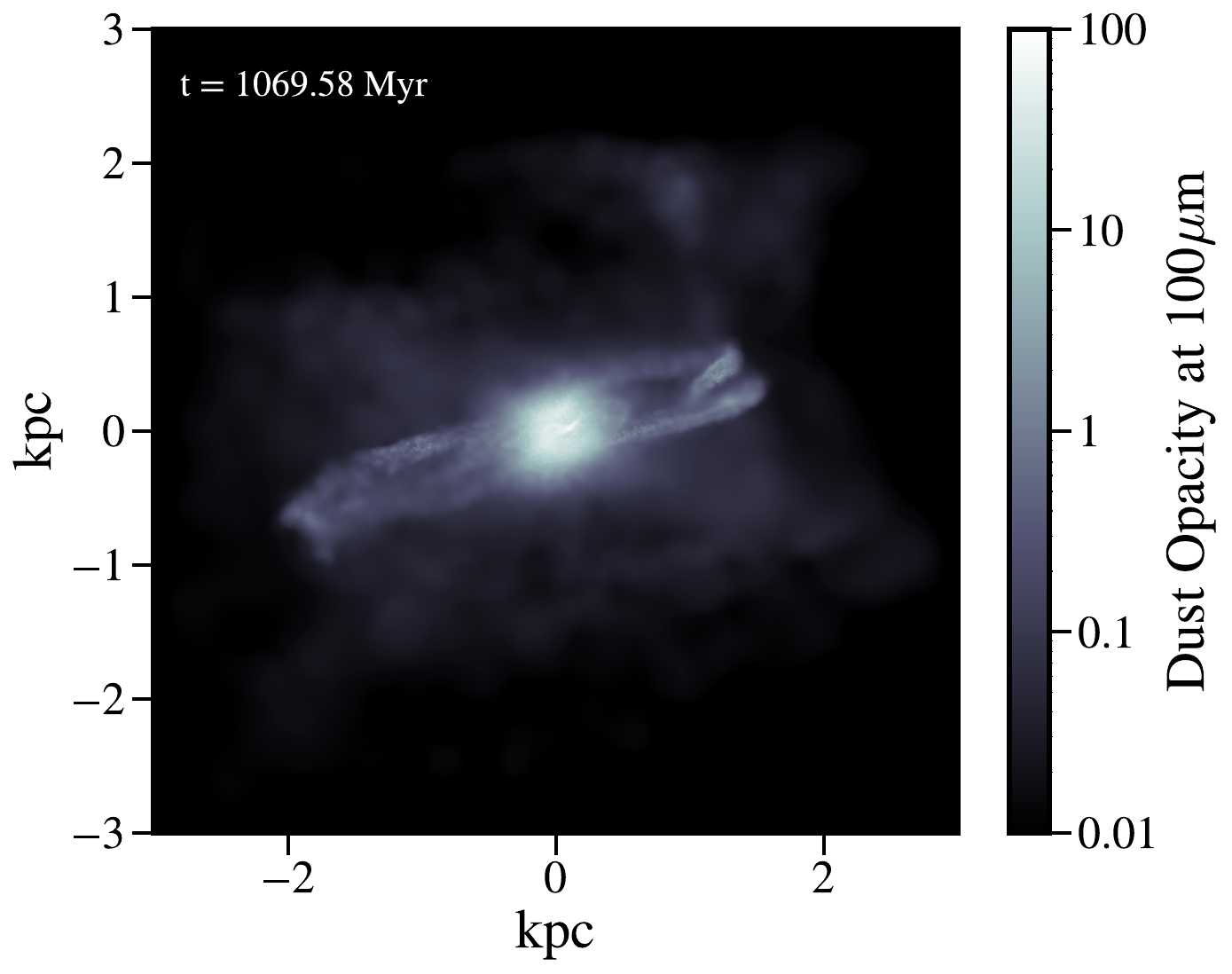}
    \caption{(Top) Dust optical depth maps at $100\mu{\rm m}$ for a galaxy at $z=4.5$ viewed face-on (left) and edge-on (right). The pixel scale in both images is \textbf{5 pc}. (Bottom) Dust optical depth maps for a different galaxy at $z=6.5$ (left) and $z=5.5$ (right).}
    \label{fig:tau_fir}
\end{figure*}

We collapse the Cosmic Sands dust temperature distributions into their luminosity-weighted values in Figure \ref{fig:lum_weighted_temp}, where we plot T$_{\rm lum-weighted}$ as a function of the total stellar mass, dust mass, and SFR surface density. The SFR surface density is calculated using the {\sc caesar} FOF defined stellar half-mass radius, computed from the radially-sorted cumulative mass of the star particles tagged by {\sc caesar} as being part of the galaxy structure. The radii range from $0.08-1.4$~kpc. The dust temperature-mass relations show that more massive galaxies have warmer dust, but there is increased scatter at the high mass end, where galaxies with M$_{*} \gtrapprox 2\times10^{10}$~M$_{\odot}$, M$_{\rm dust} \gtrapprox 8\times10^7$~M$_{\odot}$ have dust temperatures that vary by 40~K. The relation with SFR surface density shows a clearer trend, implying that luminosity-weighted dust temperature is more sensitive to density than to the integrated quantities; the high density starburst galaxies host higher surface density and warmer dust reservoirs which increase the luminosity-weighted temperatures. 

In Figure \ref{fig:Tmw_Tlw}, we compare the mass-weighted, luminosity-weighted, and peak dust temperatures. We find that luminosity-weighted temperatures are warmer than the mass-weighted, indicating that while the most luminous dust is the warm dust, most of the dust mass is colder, as expected. The ratio of T$_{\rm lum-weighted}$/T$_{\rm mass-weighted}$ ranges from $1.6-2.1$. The mass-weighted temperatures tend to increase with increasing dust mass. Also demonstrated in Figure \ref{fig:Tmw_Tlw} is the (lack of) correlation between the apparent temperature, T$_{\rm peak}$, and T$_{\rm lum-weighted}$ and T$_{\rm mass-weighted}$. For galaxies with less dust mass, the luminosity-weighted and peak temperatures agree relatively well, with some scatter driven by the temperature distributions unique to each galaxy. But the peak temperatures are much lower compared to T$_{\rm lum-weighted}$ for the more dust rich systems, due to the complex interplay between local dust temperature and emissivity properties as will be discussed in the following section. Thus, while the dust temperature distribution of a single galaxy can be represented by a single characteristic temperature, this temperature is 1) not unique to that temperature distribution and 2) not always equivalent to the other characteristic temperatures as each temperature is sensitive to a specific part of the entire dust temperature distribution.

\subsection{The Impact of FIR Optical Depth}\label{sec:optical_depth}

As shown in Figure \ref{fig:Tmw_Tlw}, the relationship between the characteristic dust temperatures (T$_{\rm lum-weighted}$ and T$_{\rm mass-weighted}$) and the apparent temperature (T$_{\rm peak}$) are complex for galaxies with large dust reservoirs, primarily due to radiative transfer effects and dust optical depth. Dust thermal emission in the FIR is typically assumed to be optically thin, or at least transitioning from optically thick to thin around $100-200\mu$m; this assumption is adopted extensively in the literature to model FIR SEDs of both UV-selected and submillimeter-selected galaxy samples \citep{swinbank_2014, scoville_2017, harikane_2020, sommovigo_2022_rebels_alpine_dust}. Here, we argue that this assumption may be invalid for dusty galaxies at high-redshift, with significant implications for the observed dust properties. 

First, in the left panel of Figure \ref{fig:temp_lir_tau}, we show the infrared luminosity of the Cosmic Sands galaxies as a function of luminosity-weighted temperature with galaxies color-coded by dust mass. The luminosities are calculated from the SEDs generated at a single random viewing angle. For galaxies with dust masses $<2\times10^7$~M$_{\odot}$, IR luminosities increase in step with increasing temperatures as is expected for optically thin emission, albeit with some scatter driven by the distribution of dust temperatures within each galaxy.\footnote{If the dust in a galaxy were at uniform temperature, we would expect the global IR luminosity to be highly correlated with the apparent temperature (and in the optically thin limit, we should expect the apparent temperature to be very close to the intrinsic luminosity-weighted temperature, as evidenced by Figure \ref{fig:temp_lir_tau}). But because in all galaxies there is an intrinsic distribution of dust temperatures, there will be some scatter in the T$_\mathrm{luminosity}$ - LIR relationship, since the FIR SED will essentially be the sum of all of the modified blackbody spectra from the dust at different temperatures.} But this correlation plateaus at larger temperatures and dust masses, implying that the increase in luminosity with increasing temperature does not occur as efficiently in these starbursty, dust rich systems and could be driven by the fact that these particular galaxies are optically thick at FIR wavelengths. 

We demonstrate this with two example galaxies in Figure \ref{fig:tau_fir}, where we show the dust optical depths for a face-on and edge-on orientation at $100\mu$m over scales of $5$~pc. The optical depths, $\tau_i = \int\kappa~\rho_i~ds$, are calculated in each {\sc powderday} cell $i$ with dust density $\rho_i$ along (the viewing angle dependent) line of sight $ds$ from $-5$ to $5$~kpc with respect to the center of mass of the galaxy, assuming absorption cross sections of $\kappa = 40.95$~cm$^2$/g at $100\mu$m \citep{draine_03_araa, li_draine_01_2, weingartner_draine_2001}. In the top panels, which shows a galaxy with M$_{\rm dust} = 1.2\times10^9$~M$_{\odot}$ at $z=4.5$, we find that when viewing the galaxy face-on, the central 0.1-0.5 kpc of the galaxy is opaque at $100 \mu$m, with $\tau \sim 25$ in the densest region. When viewed edge-on, the thin disk has $\tau \sim 30$, with the thick disk marginally opaque spanning $\tau \sim 1-5$. In the bottom panel a galaxy at $z=6.5$ (left) and $z=5.5$ (right) with a dust mass of M$_{\rm dust} \sim 6\times10^8$~M$_{\odot}$ has dust opacities of $50-100$ in the central 10-100 pc. 

Similarly high dust opacities are found in compact-obscured nuclei \citep[CONs][]{aalto_2019, falstad_2021} and ULIRG systems with CONs at low-z including Arp 220, whose FIR/submillimeter emission does not become optically thin until $\sim2.5$mm \citep{scoville_2017}. The spatial extent of these high dust column densities vary depending on the source: the compact nucleus of the nearby IR luminous galaxy IC860 is opaque at millimeter wavelengths on scales of $<10$ pc \citep{aalto_2019} while the disks of Arp 220 are opaque at $100\mu$m ($2.6$mm) on scales of $240$ pc ($100$ pc) \citep{rangwala_2011, scoville_2017}, comparable to the highest optical depths of the Comic Sands galaxies at $500$ pc.

At high redshift, interferometric imaging of IR luminous galaxies have revealed opaque structures on kpc scales. For instance, SMA and LABOCA imaging of the Cosmic Eyelash \citep[SMMJ2135-0102,][]{swinbank_2010} resolved optically thick clumps embedded in the more diffuse ISM, with sizes estimated at $\sim500$ pc; the average ISM column density of the system on the whole is estimated at $\sim10^{24}$ cm$^{-2}$ over a $1$ kpc radius \citep{danielson_2011}. Similarly, ALMA detections of [CII] in the SPT SMG 0311-58 resolved clumpy structure on scales of $0.6-1.3$ kpc with [CII] emission suggesting high column densities of gas $(0.8-12 \times 10^4$ M$_{\odot}$ kpc$^{-2})$ \citep{spilker_2022}. Lastly, dust properties inferred from NOEMA and ALMA detections of CO in several compact high-z starburst galaxies favored optically thick $100\mu$m emission ($\tau \sim 1.5-4$) over kpc scales \citep{jin_2022}. 

Explicit predictions for large FIR dust opacities from hydrodynamical simulations are sparse, but \cite{yajima_2014} report UV extinction towards individual stars in $z=3$ Lyman-break galaxies modeled with {\sc gadget-3}. For some stars in the most massive halos, they calculate optical depths in excess of $100-1000$ at $1700\AA$, assuming the graphite and silicate dust model of \cite{draine_lee_1984}. Though the pencil beam estimates do not compare spatially to the values calculated for the Cosmic Sands galaxies, some sight-lines towards the \cite{yajima_2014} halos could be opaque in the FIR if the UV extinction is $>1000$. Similarly, \cite{lovell_2022_orientation} estimate the UV extinction along individual sight-lines towards galaxies at $z=2$ from the $100$~Mpc {\sc simba} simulation. They find (using the column density of dust as a proxy for the dust emission along that line of sight) that some galaxies have UV opacities in excess of $1000$, again implying that these systems could also be opaque in the FIR. All together, these findings suggest that while the column densities of the Cosmic Sands galaxies may be extreme, they are comparable to other dusty, star-forming systems at both low and high redshift.

\begin{figure*}[t]
    \centering
    \includegraphics[width=0.9\textwidth]{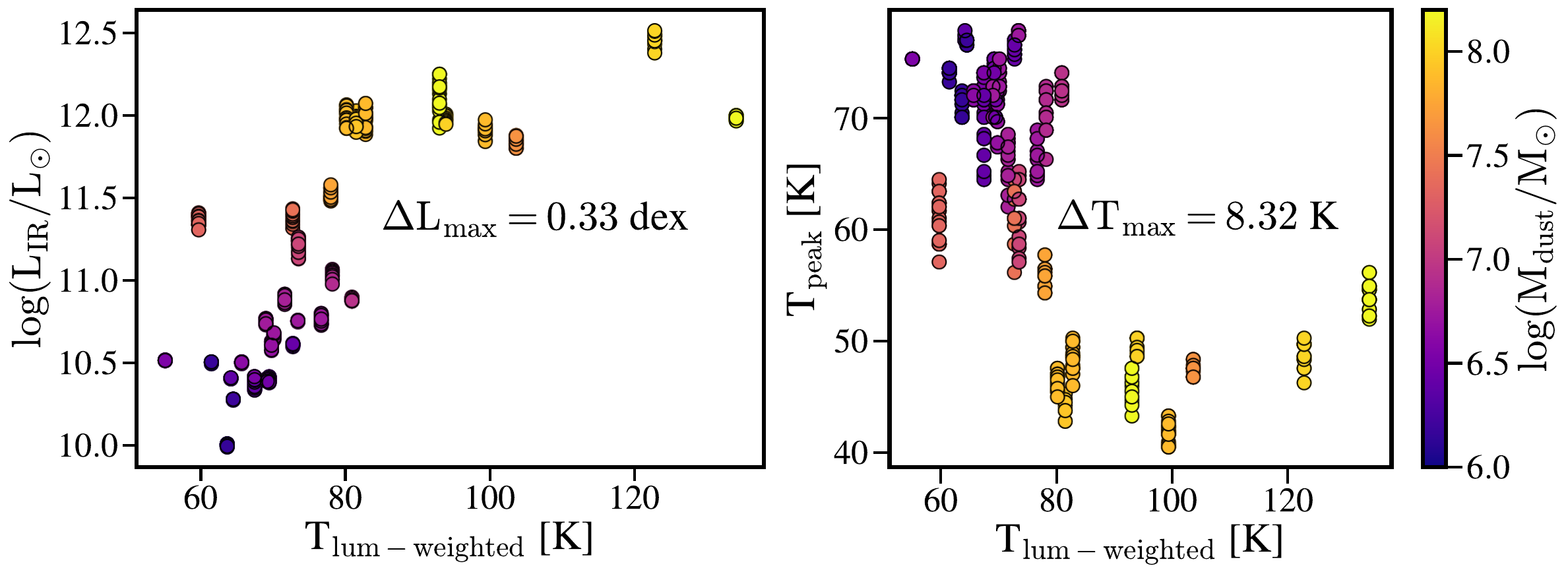}
    \caption{Infrared luminosity (left) and temperature corresponding to the peak for the FIR SED (right) as a function of luminosity-weighted dust temperature. Points are color-coded by galaxy dust mass. Each Cosmic Sands galaxy is represented by $25$ points, corresponding to $25$ SEDs generated at different viewing angles, showing the dependence of IR luminosity and apparent temperature on the orientation angle of the galaxy. The maximum spread in luminosity (0.33 dex) and temperature (8 K) are shown in each panel.}
    \label{fig:Tpeak_distr}
\end{figure*}

The consequence of these high FIR dust opacities is that observationally, only the dust at the $\tau \lesssim 1$ surface is detected, which can bias the measured dust luminosity, temperature, and mass of these systems if the optical depth is not accounted for \citep[e.g.][]{cortzen_2020, jin_2022, fanciullo_2020}. Additionally, these large dust opacities in the Cosmic Sands galaxies explain the slowing trend between the luminosity and temperature for the most dust rich systems in left panel of Figure \ref{fig:temp_lir_tau}: the hottest dust in these galaxies is obscured by cooler dust in the foreground from which the IR photons we measure originate from. Returning to Figure \ref{fig:temp_lir_tau}, we show this explicitly in the right panel, where we plot IR luminosity versus T$_{\rm lum-weighted}$ but this time color-coded by the maximum dust optical depth at $100\mu$m calculated in cells of $0.1$~kpc\footnote{We note that optical depth is a line-of-sight quantity that varies spatially within a single galaxy and that, like the dust temperature, quantifying the `average' dust optical depth for a galaxy is nontrivial; for instance, assuming optical depth varies as a powerlaw function of frequency (as in the case of a modified blackbody, $\tau \propto \nu^{\beta}$), we do not get reasonable fits to the FIR SEDs of the Cosmic Sands galaxies, and so we choose to simply characterize our galaxies by their maximum dust optical depth averaged over $0.1$~kpc scales. Also note that the optical depths reported in Figure 6 are calculated on smaller scales than the optical depths reported elsewhere in the text.}. The galaxies whose luminosities are not as high as their luminosity-weighted temperatures would suggest have dust opacities $>1$, implying that we cannot see emission from the densest, hottest dust. In combination with Figure \ref{fig:lum_weighted_temp}, higher density starburst galaxies have more intense UV/optical radiation fields and have warmer, higher density dust with higher optical depths. As we'll explore in the following sections, one of the major simplifying assumptions made when fitting a modified blackbody function to galaxy SEDs is that the emission is optically thin at long wavelengths, which could severely limit the ability to accurately infer dust temperatures in the most dust rich systems.

\subsection{Orientation Angle Bias}

For both observed galaxies \citep[e.g.][]{nagaraj_2022} and galaxies from hydrodynamical simulations \citep[e.g.][]{lovell_2022_orientation}, it has been shown that variations in measured galaxy properties (e.g., luminosity, UV-NIR dust attenuation) are significantly impacted by the galaxy's morphology and orientation with respect to the observer. We briefly make note of this in the context of biases in measuring dust properties from FIR SEDs.

\begin{figure*}[t!]
    \includegraphics[width=\textwidth]{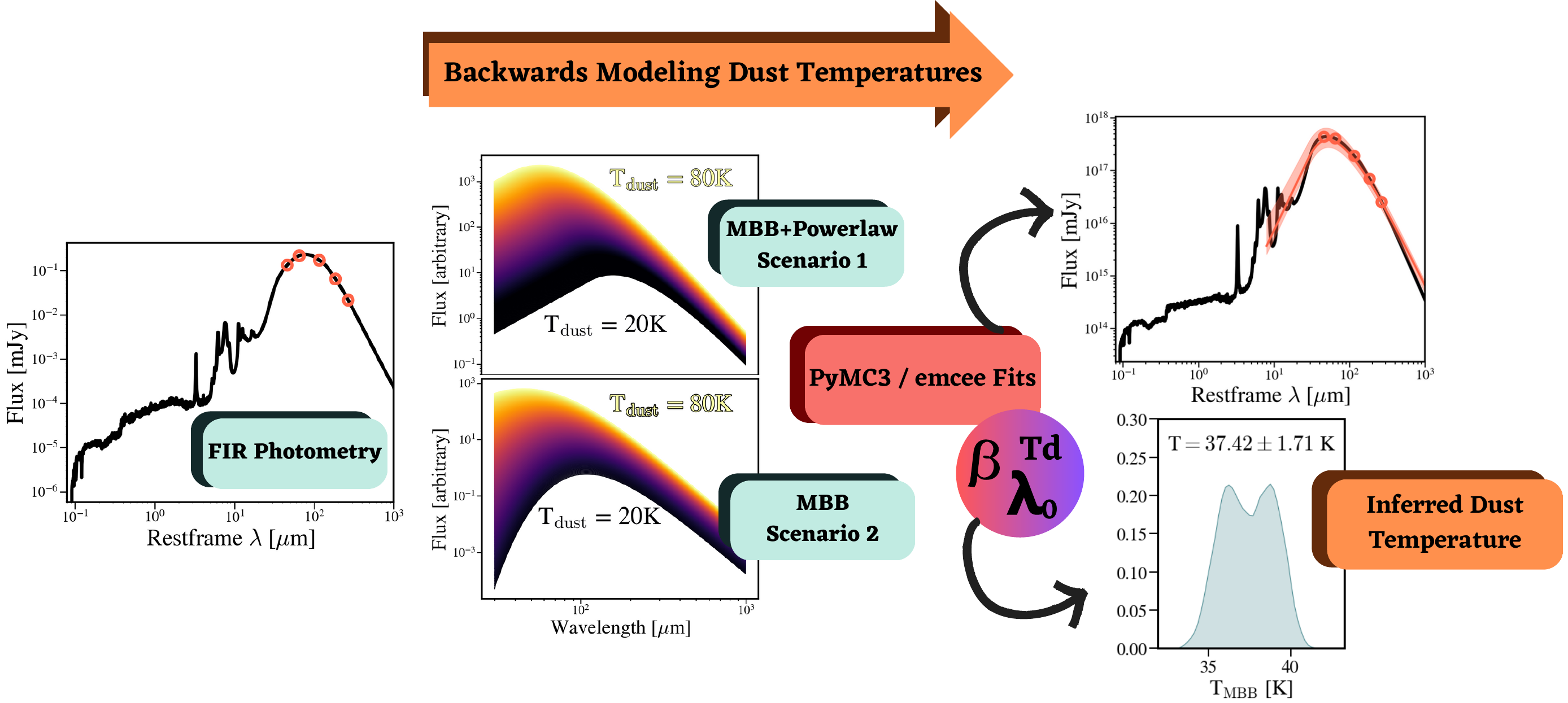}
    \caption{Schematic detailing the backwards dust temperature modeling with modified blackbody fitting. We first sample photometry from the {\sc powderday} SED for each galaxy (5 bands for Scenario 1 and 2 bands for Scenario 2). For Scenario 1, we fit a modified blackbody function (MBB) with a powerlaw component on the Wien's side of the FIR SED. For Scenario 2, we fit just a modified blackbody function since the two photometric constraints will not constrain the powerlaw on the Wien's side. The parameter posteriors are sampled with {\sc pymc3} or {\sc emcee} from which we infer the dust temperature for each galaxy.}
    \label{fig:mbb_diagram}
\end{figure*}

To demonstrate this, we generated SEDs at $25$ different viewing angles for each Cosmic Sands galaxies. Figure \ref{fig:Tpeak_distr} now shows the dependence of the IR luminosities and peak temperatures on the orientation angle of each galaxy. The maximum spread in luminosity (0.33 dex) and temperature (8 K) are shown in each panel. What these plots demonstrate is the anisotropic nature of the IR emission and assuming isotropic emission neglects the effects of star-dust geometry and galaxy morphology. The consequences of this for observations of real galaxies is that the IR luminosity and dust temperatures measured from galaxy SEDs are subject to an uncertainty that is non-trivial to quantify. 

\section{Backwards Modeling Dust Temperature}\label{sec:measuring_dust_temps}

We now approach modeling dust temperatures from the opposite side: inferring dust properties from galaxy SEDs. In practice, this is a relatively challenging task, given the degeneracies between FIR luminosity, dust temperature, and dust mass that drive variations in the shape of the FIR SED which can be difficult to constrain with limited data. As such, several approaches to determine dust temperatures exist in this regime, all of which depend on the assumed shape of the FIR SED and dust optical properties from the UV-FIR. We focus our analysis on the most common method for fitting FIR SEDs of galaxies at both low and high redshift: fitting a modified blackbody function to broadband FIR data\footnote{Of course, there are other models that have been successful at modeling thermal dust emission on both molecular cloud scales and galactic scales, including the variable interstellar radiation field (ISRF) models of \cite{dale_2002} and \cite{draine_infrared_2007} which effectively allow for a distribution of dust temperatures due to the differential heating of dust grains. But, like the modified blackbody model, the dust properties inferred from these models are dependent on the model assumptions, namely, the nature of the ISRFs and the chosen dust grain properties.}.

\subsection{Model Definitions}

If we assume modified blackbody emission from a source with a single temperature \citep{hildebrand_1983}, T$_{\rm MBB}$ is the dust temperature that fits the SED in the following, neglecting any distance/redshift terms:
\begin{equation}
    {\rm S}_{\nu} \propto {\rm B}_{\nu}({\rm T}_{\rm MBB})~(1 - e^{-\tau_{\nu}})~\epsilon_{\nu},
\label{eq:mbb_prop}
\end{equation}
where ${\rm B}_{\nu}$ is the Planck function, $\tau_{\nu}$ is the frequency-dependent optical depth, and $\epsilon_{\nu}$ is the emissivity, which accounts for the `imperfect' blackbody (greybody) nature of dust grain emission in the FIR. We can simplify this equation since $\tau$ is proportional to $\epsilon$ and assume the typical frequency-dependent powerlaw parameterization for $\tau$ to get\footnote{Given the nature of dust thermal emission, the absorption and emission of dust grains are intrinsically linked; the emissivity is proportional to the opacity which is proportional to the optical depth (and are sometimes used interchangeably in the literature), all of which can be described by a frequency-dependent powerlaw function in the FIR, with a variety of normalization factors used in the literature \citep[e.g.][]{kovacs_2010, reuter_2020, algera_2023_new_rebels} from fits to observations of galaxies, clouds, and disks at different wavelengths and redshifts. The foundational work of \cite{hildebrand_1983} defined a dimensionless emissivity $Q_{\nu}$ that depends on grain size, geometry, and the frequency-dependent mass-absorption coefficient $\kappa_{\nu}$ (opacity), describing the efficiency of emission from the grain surface area per unit mass. Like $\beta$ with dust temperature, the assumed value of $\kappa_{\nu}$ significantly impacts the dust masses inferred from FIR SEDs.}:
\begin{equation}
    {\rm S}_{\nu} \propto {\rm B}_{\nu}({\rm T}_{\rm MBB})~(1 - e^{(\nu/\nu_0)^{\beta}}),
\label{eq:mbb_beta}
\end{equation}
where the optical depth is defined in such a way to be unity at $\nu_0$ and the powerlaw slope is set by $\beta$, the spectral emissivity index. For an individual grain of known size and composition, the emissivity can be measured but in practice the emissivity of the entire dust population is difficult to determine due to dependencies on the dust temperature distribution and line-of-sight column density. The circular dependence of $\beta$ on the local dust properties (temperature, grain size and composition) and of T$_\mathrm{MBB}$ on $\beta$ when fitting SEDs alludes one of the primary difficulties in constraining dust temperatures.

Because a modified blackbody fit alone does not typically describe the mid-IR excess found in the Wien's side of the thermal emission peak due to the presence of warm dust, the short wavelength FIR data ($\lambda < 50 \mu$m) is typically fit with a power law:
\begin{equation}
{\rm S}_{\nu} \propto (1 - e^{(\nu/\nu_0)^{\beta}})~{\rm B}_{\nu}({\rm T}_{\rm MBB}) + \lambda^{\alpha} e^{-(\lambda / \lambda_c)^2},
\label{eq:mbb_powerlaw}
\end{equation}
where $\alpha$ is the slope of the MIR powerlaw component and $\lambda_c$ is the wavelength where the SED transitions from powerlaw dominated to modified blackbody dominated. In both cases (Equation \ref{eq:mbb_beta} MBB or Equation \ref{eq:mbb_powerlaw} MBB$+$powerlaw), T$_{\rm MBB}$ will correspond to the temperature of the dust at the $\tau \sim 1$ surface.

\begin{figure*}[t!]
    \centering
    \includegraphics[width=0.98\textwidth]{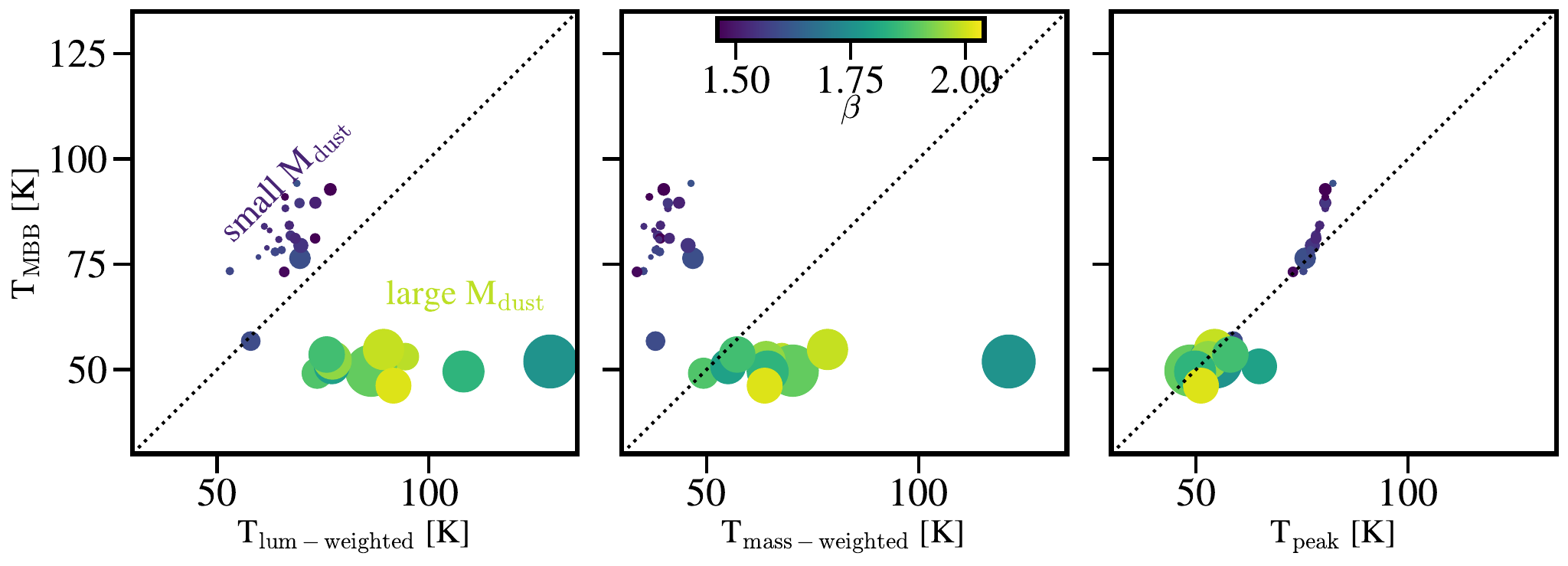}
    \caption{Dust temperature inferred from modified blackbody+powerlaw fits to the Cosmic Sands SEDs as a function of luminosity-weighted temperature (left), mass-weighted temperature (middle), and temperature corresponding to the peak of the FIR SED (right). Points are color-coded by the median of the emissivity spectral index ($\beta$) posterior from the modified blackbody+powerlaw fit. The sizes of the points reflect their dust mass, where galaxies with larger dust masses have larger markers.}
    \label{fig:TBB_true_temps}
\end{figure*}

The spatial variation of the underlying distribution of dust temperatures in a galaxy can necessitate the need for additional flexibility that a single-temperature modified blackbody model does not allow; as shown in Figure \ref{fig:temp_dist}, the temperature distributions could be most accurately described by multiple temperature components (e.g., cold, warm, hot) that dominate the emission at different wavelengths. For galaxies with well constrained FIR SEDs, the best route would be to fit the data with a multi-temperature component model \citep{casey_2012} instead of the more commonly used single temperature modified blackbody model above. But as demonstrated by \cite{rangwala_2011}, while the single temperature model neglects the details of the spatial variation of dust temperatures within a galaxy, attempts to use a two-temperature optically thin model to fit SEDs of an extremely optically thick galaxy can result in erroneous estimates of other dust properties (e.g., dust masses that do not align with estimates from molecular gas constraints even though the inferred dust temperatures are physically reasonable). Thus, it is also important to understand the degeneracies between multiple temperature components and optical depth as one can mimic the other's SED fingerprints \citep{dunne_2001}. In 

We note here that the typical use of Equation \ref{eq:mbb_beta} in the literature is to constrain the dust temperature, which can then be used to measure the dust mass. However, as noted in \cite{casey_2012} and demonstrated in \cite{cochrane_2022}, the dust temperature has a profound effect on the inferred dust mass when not in the RJ regime of the FIR SED: since the Planck function is now proportional to $(e^{h \nu / k {\rm T}_{\rm MBB}} - 1)^{-1}$, a $5$~K difference in measured T$_{\rm MBB}$ (from $30$~K to $35$~K) results in a $\sim125$\% increase in the inferred dust mass.

\begin{figure*}
    \centering
    \includegraphics[width=0.47\textwidth]{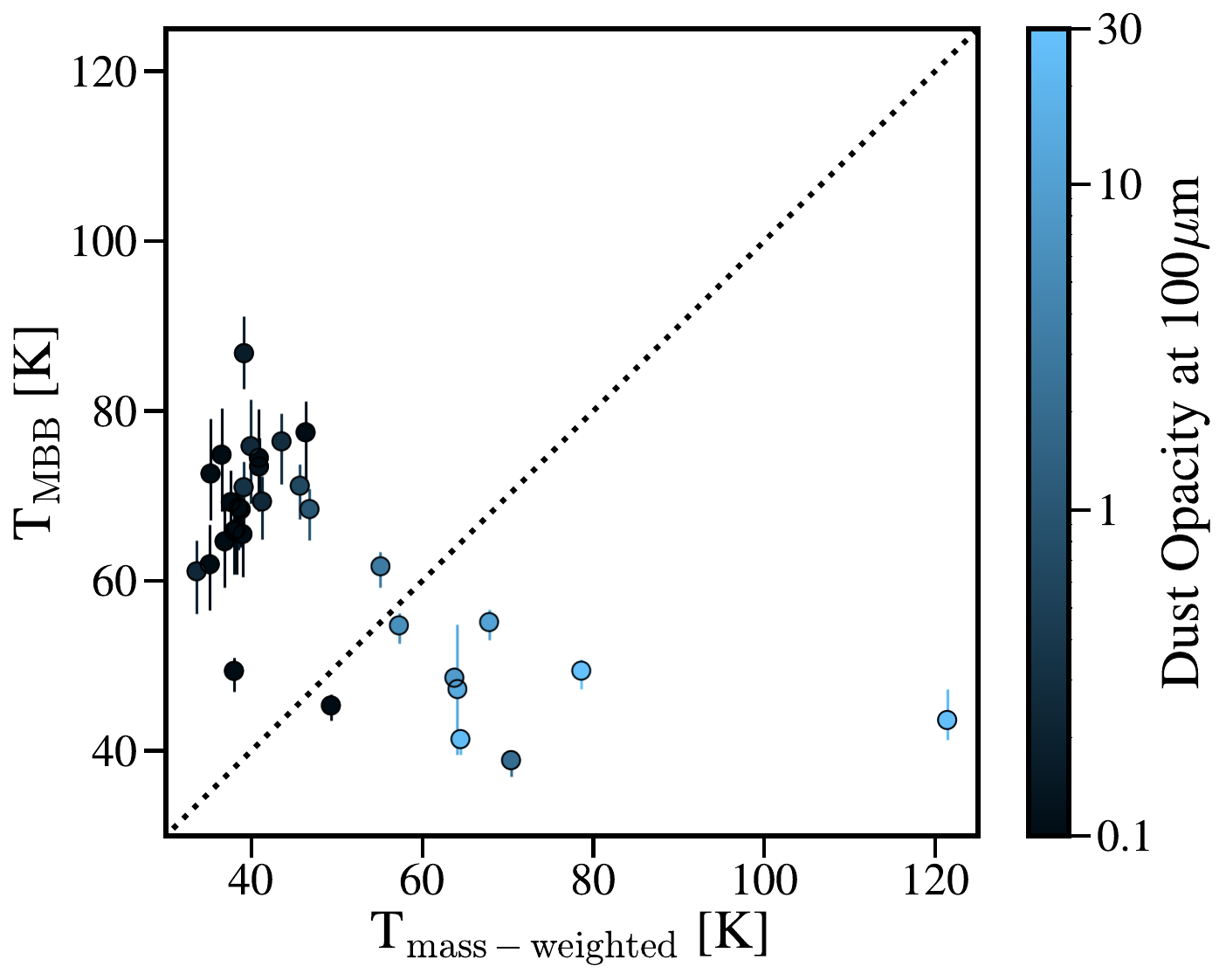}
    \includegraphics[width=0.47\textwidth]{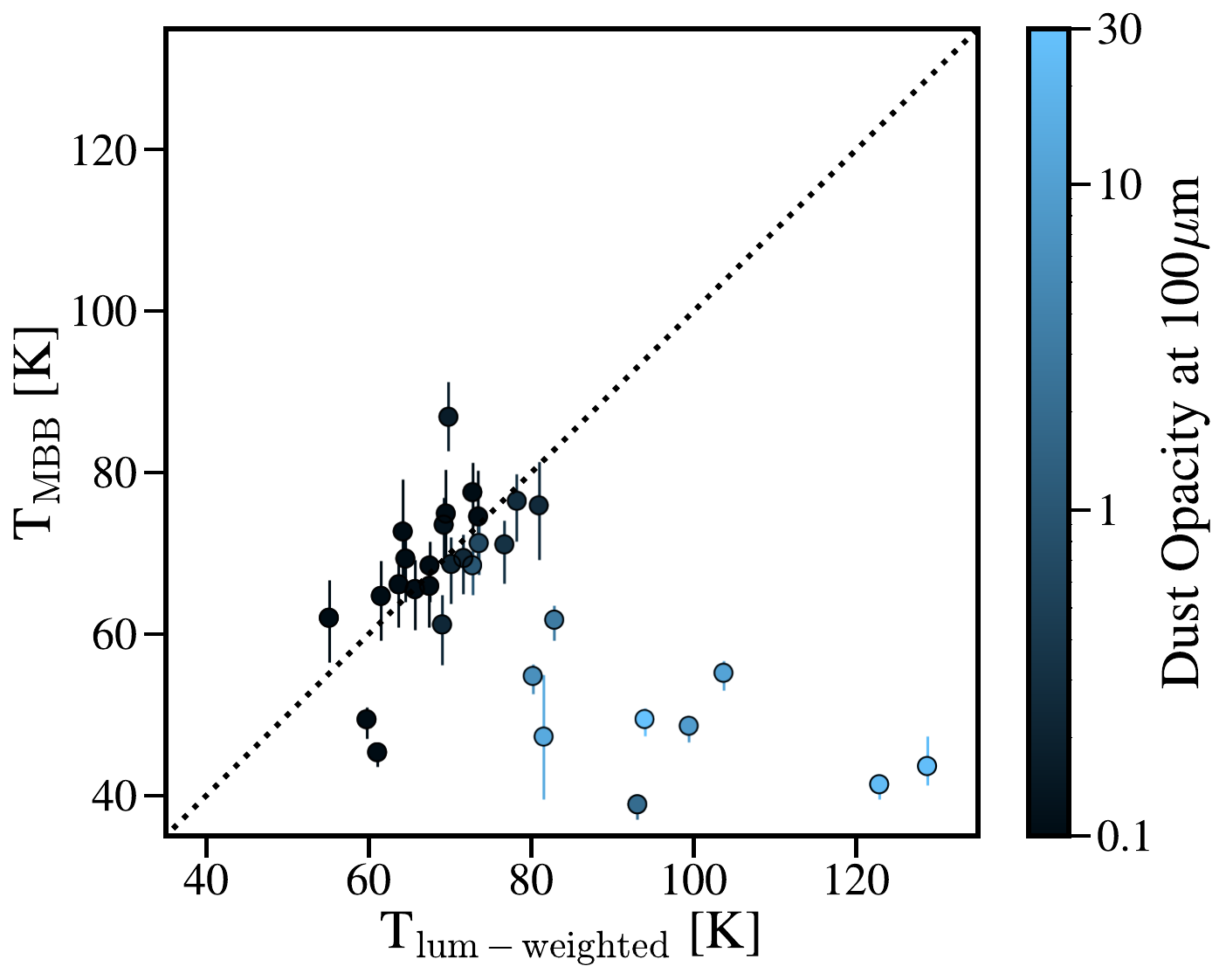}
    \caption{Dust temperatures inferred from modified blackbody fits described in Scenario 1 as a function of mass-weighted (left) and luminosity-weighted (right) temperature. Points are color-coded by FIR optical depth at $100\mu$m as calculated in \S \ref{sec:optical_depth}.}
    \label{fig:Tbb_full}
\end{figure*}

\subsection{Characterizing the Cosmic Sands FIR SEDs}\label{sec:characterizing_sed}

In Figure \ref{fig:mbb_diagram}, we show the process of inferring the Cosmic Sands dust temperatures from modified blackbody fits to photometry sampled from the {\sc powderday} SEDs. The basic idea is to treat the Cosmic Sands galaxies and their SEDs as though we are `observing' them, and test the efficacy of modified blackbody fitting on high-z dusty galaxies. First, however, to understand how well the Cosmic Sands galaxies are characterized by a modified blackbody function, we fit Equation \ref{eq:mbb_powerlaw} to the full {\sc powderday} SEDs from $30-1000\mu$m\footnote{We randomly choose one of $25$ SEDs generated at different viewing angles for each galaxy to fit; though this will not change the true mass-weighted and luminosity-weighted temperatures, as shown in Figure \ref{fig:Tpeak_distr}, this can impact the apparent dust temperatures and the shape of the FIR SED.}. We use the fitting routine {\sc mcirsed} introduced in \cite{drew_2022} and allow all parameters (T${_\mathrm{MBB}}$, $\beta$, $\nu_0$, and $\alpha$) to vary. {\sc mcirsed} uses the {\sc PyMC3} package \citep{salvatier_2016} to sample parameter posteriors, which are constrained by uniform priors spanning T$_{\rm CMB}(z=6.5) <$ T$_{\rm MBB} < 250$~K, $0.5 < \beta < 3.0$, $0.5 < \alpha < 4.0$, and $30 < \lambda_0 < 400\mu$m.

In Figure \ref{fig:TBB_true_temps}, we plot T${_\mathrm{MBB}}$ as a function of the true dust temperatures shown in Figure \ref{fig:temp_dist}: luminosity-weighted, mass-weighted, and the temperature corresponding to the peak wavelength of the FIR SED. The points are color-coded by the median of the $\beta$ posterior distribution and the sizes of the points correspond to the dust mass of each galaxy, with larger points indicating larger dust masses. First, we note that the Cosmic Sands SEDs span a range of $\beta$ values but are generally concentrated around $\beta=1.5$ and $\beta=1.9$, with the dust rich systems favoring larger $\beta$ values. Recall that as part of the forward modeling process, the dust emission from each particle is a function of the particle temperature, assuming modified blackbody emission with the emissivity in turn dependent on the \cite{weingartner_draine_2001} grain properties. Each particle containing dust can theoretically emit a blackbody with a unique emissivity, resulting in an effective FIR SED on galactic scales made up of various blackbody spectra at different temperatures and emissivities. The observed effective emissivity is then dependent on the emissivity of each particle and the column density of the dust, which will impact the optical depth towards the observer. What this means is that from the observed SED, we can really only probe this effective emissivity and optical depth, which are both dependent on the underlying temperature distribution. A consequence of this is that an apparently cold system with a measured emissivity of $\beta = 2$ can either be due to optically thin emission from cold dust or optically thick cold dust obscuring a temperature gradient towards warm dust \citep{jin_2022}.

Unsurprisingly, the modified blackbody temperatures are generally a poor estimate for both the luminosity-weighted and the mass-weighted dust temperatures. For dust rich systems, the blackbody fit under-estimates both temperatures because the dust column density obscures the rest of the warmer dust distribution from view, thereby appearing colder. On the other hand, the galaxies with lower dust content have blackbody temperatures that are too warm compared to their physically weighted temperatures. Because the optical depth in these systems is much smaller, we can ``see" into the entire temperature distribution, but the SED will be dominated by the emission from the warmest dust. Thus, while still biased warm, the blackbody temperatures are a bit closer to the luminosity-weighted temperatures than the mass-weighted. For both high and low dust mass galaxies, the blackbody temperatures are closest to the temperatures corresponding to the peak of the FIR SED as they are both measures of apparent temperature. However, we note that there is an average bias of $6.25$~K which is smaller than the offsets from the physical temperatures but relatively large compared to the dynamic range of the effective dust temperatures and the uncertainties associated with the fit.

\begin{figure*}
    \centering
    \includegraphics[width=0.47\textwidth]{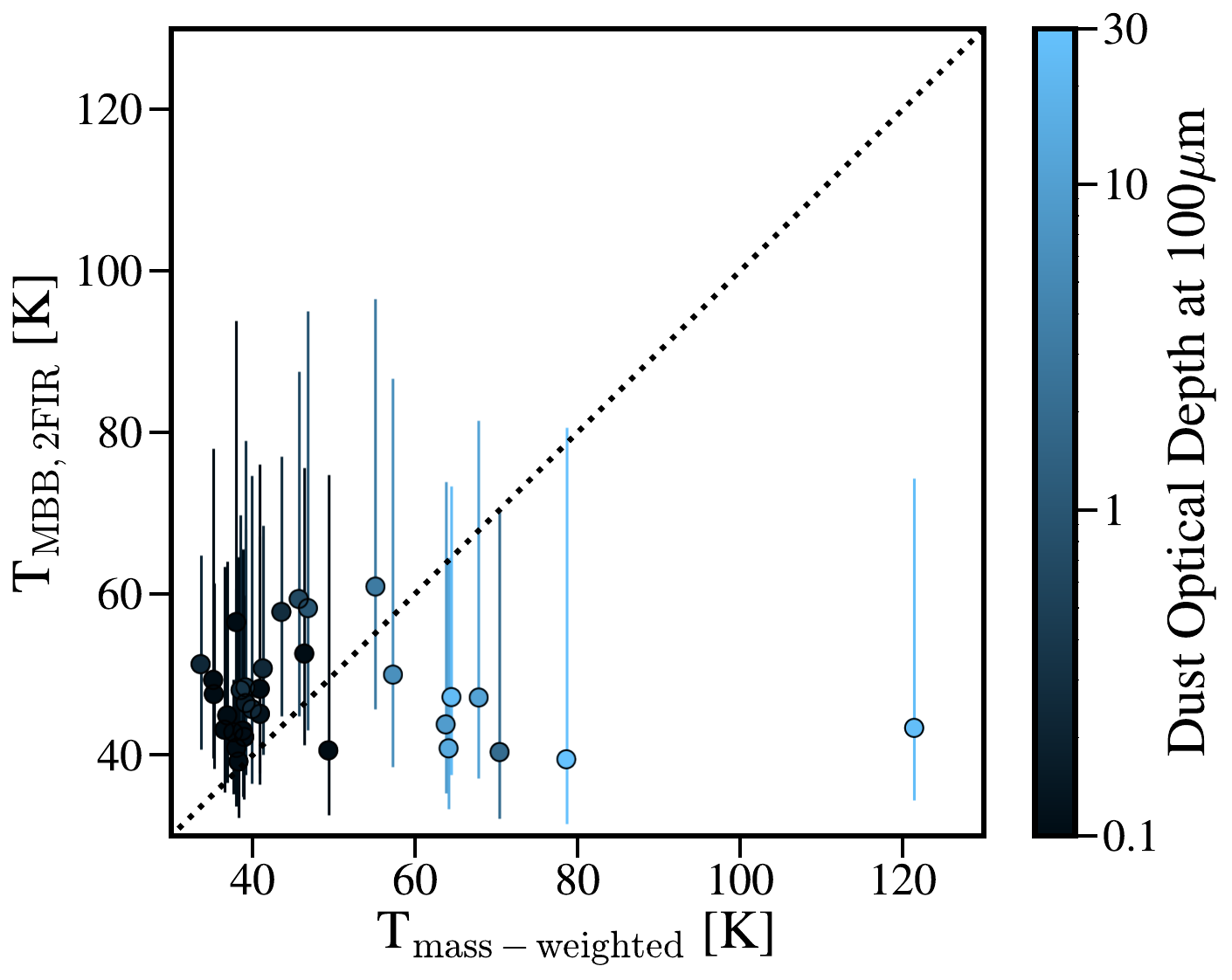}
    \includegraphics[width=0.47\textwidth]{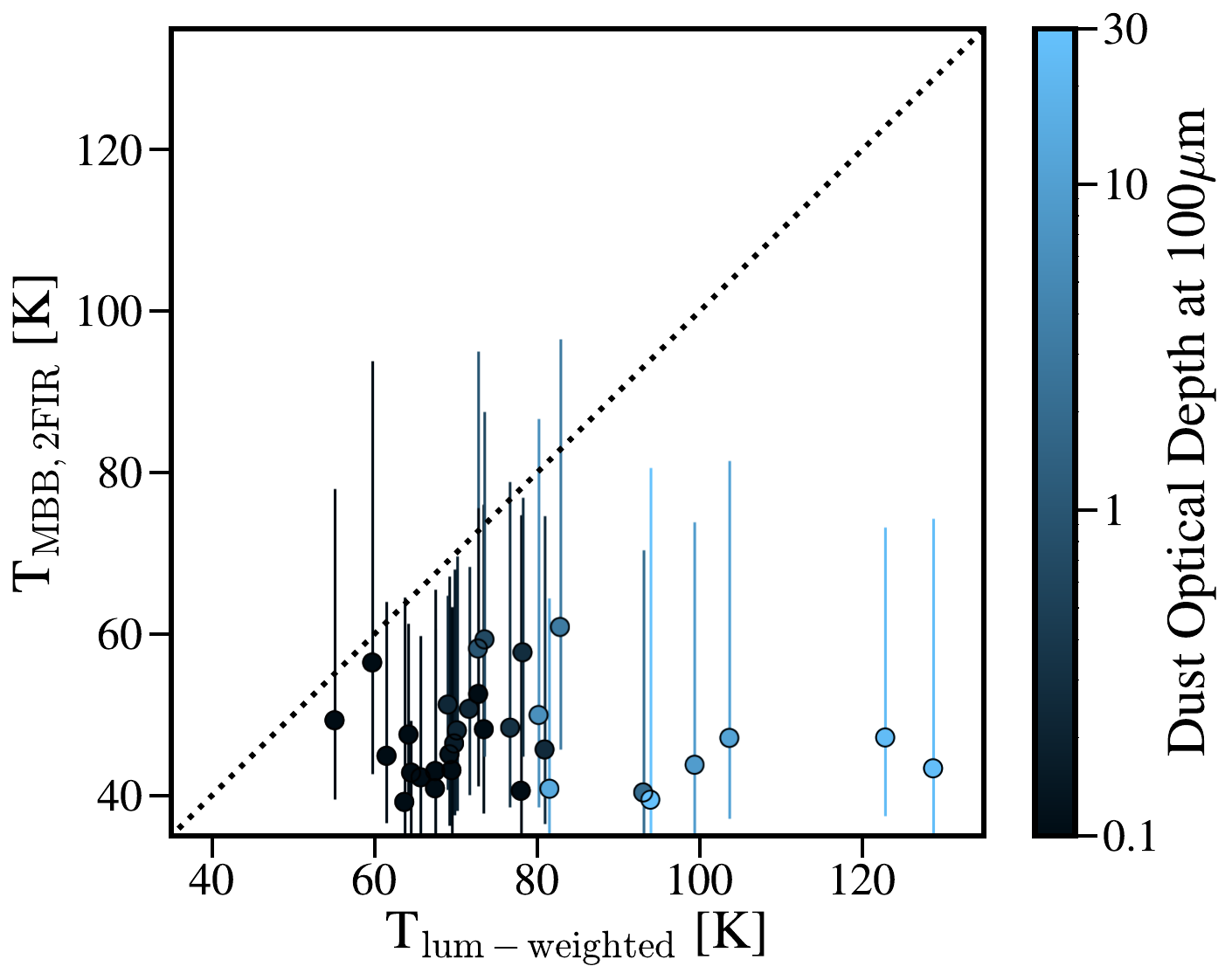}
    \caption{Dust temperatures inferred from modified blackbody fits described in Scenario 2 as a function of mass-weighted (left) and luminosity-weighted (right) temperatures. Points are color-coded by FIR optical depth at $100\mu$m as calculated in \S \ref{sec:optical_depth}.}
    \label{fig:Tbb_2}
\end{figure*}

\subsection{Fitting FIR Photometry with a Modified Blackbody Model}\label{sec:fitting_fir_phot}

Fitting a modified blackbody+powerlaw function to the Cosmic Sands SEDs generally produces reasonable SED models, but even in a `best-case scenario' in which the true shape FIR SED is known, the accuracy of the measured dust temperatures depends on the dust content of the galaxy, specifically the dust optical depth in the FIR. The primary challenges in inferring dust properties from SEDs are the typically sparse FIR SED coverage and the degeneracies between both the intrinsic dust properties and the chosen model parameters that are only marginally broken with shorter wavelength FIR constraints. To explore how these factors contribute to modeling dust temperatures, we test two different scenarios that differ in SED coverage using the $z=6.5$ Cosmic Sands galaxy sample. 

\subsubsection{Scenario 1: Well-Sampled FIR SED}

First, we assume optimal SED coverage and fit the modified blackbody + MIR powerlaw model in Equation \ref{eq:mbb_powerlaw} to five FIR bands (rest-frame $\lambda_{\rm rest} \sim 46, 65, 116, 185, 267 \mu{\rm m}$ which correspond to observations at 350, 500, 870, 1400, and 2000 $\mu$m for a z=6.5 galaxy). We again use {\sc mcirsed} to sample parameter posteriors. While the effective $\beta$ of the Cosmic Sands galaxies is a bimodal distribution, the intent of this exercise is to show the impact of balancing the number of free parameters with available SED constraints. Thus, we fix $\beta = 1.5$ and $\alpha=2.0$ but allow $\nu_0 = c / \lambda_0$ to vary, enabling optically thick solutions towards $\lambda_0\sim400\mu$m in the rest-frame. The parameter posteriors are constrained by a uniform priors spanning T$_{\rm CMB}(z=6.5) <$ T$_{\rm MBB} < 250$~K and $30 < \lambda_0 < 400\mu$m. We note that the model and parameter choices are similar to analysis of high-z dusty galaxies with reasonable FIR coverage that have stellar and dust properties comparable to the Cosmic Sands galaxies \citep[e.g.][]{greve_2012, spilker_2016, lower_2022_cosmic_sands} and can therefore serve as illustrative comparisons. 

We plot the median and $16^{th}-84^{th}$ percentile width of the dust temperature posteriors for each galaxy in Figure \ref{fig:Tbb_full}, comparing the the mass-weighted temperature (left panel) and the luminosity-weighted temperature (right panel). The results from these fits are nearly identical to the baseline fits to the entire FIR SED: for galaxies with low dust opacities, the dust temperatures inferred from modified blackbody fits to a well sampled SED over-estimate the mass-weighted temperatures but are reasonable estimates for the luminosity-weighted temperatures, with an average bias of $7.3$~K for galaxies with $\tau_{100\mu{\rm m}} < 0.5$. In these galaxies, the luminosity of the FIR SED is dominated by warm dust which is probed by the modified blackbody fits, given the ability to properly model the transition from optically thick to thin emission. Also similar to Figure \ref{fig:TBB_true_temps}, the dust temperature estimates for the most optically thick galaxies tend to under-estimate both the mass-weighted and the luminosity-weighted temperatures, as the bulk of the warm-hot dust mass is still highly obscured by colder dust. From this, we note the importance of allowing for general opacity solutions but also caution the utility of single temperature fits to determine the underlying dust properties as even with a relatively well sampled FIR SED, the apparent temperatures do not necessarily align with physically meaningful temperatures.

\subsubsection{Scenario 2: Sparsely Sampled FIR SED}

Imitating large FIR surveys of high-z galaxies that typically have sparse wavelength converge, we fit the modified blackbody model in Equation \ref{eq:mbb_prop} to two FIR bands (rest-frame $\lambda = 100, 160 \mu{\rm m}$) which allow us to neglect the powerlaw component that would affect data blue-ward of $\sim50\mu$m. As in the first scenario, we fix $\beta = 1.5$ and assume that emission approaches optically thick at $\lambda_0 = 100 \mu{\rm m}$. We use the {\sc emcee} package \citep{emcee} to sample the dust temperature posteriors, which are constrained by a uniform prior spanning T$_{\rm CMB}(z=6.5) <$ T$_{\rm MBB} < 250$~K. Here, our mock data and model choices are similar to analysis from \cite{algera_2023_new_rebels}, and can again serve as illustrative comparisons for dust temperature measurements from sparse data. 
    
 We show the modified blackbody dust temperatures as a function of mass-weighted and luminosity-weighted temperatures in the left and right panels of Figure \ref{fig:Tbb_2}, where we plot the median and $16^{th}-84^{th}$ percentile width of the T$_{\rm MBB}$ posterior for each galaxy. Though this model accounts for FIR dust optical depth, the wavelength at which dust transitions from optically thick to thin is fixed at $100\mu$m. First, the uncertainties on the dust temperatures inferred from these fits are much larger than the previous model, with an average posterior posterior width of $35.5$~K, similar to the uncertainties in dust temperature measurements in \cite{algera_2023_new_rebels} across their 3 model variants. This is mainly driven by the fact that only $2$ photometric bands are being used to constrain $2$ model parameters (amplitude, temperature) and that while the shortest wavelength constraint at $100\mu$m is nearer to the peak of emission, it is still not a strong enough lever arm to break the degeneracies between the wide parameter space that fits the Rayleigh-Jeans tail of these SEDs. 

If we focus on the median inferred temperatures, these are marginal over-estimates of the mass-weighted temperatures of less dusty galaxies and marginal under-estimates of the luminosity weighted temperatures. The dust-rich galaxies, on the other hand, all have similar temperature estimates ($40-50$~K). As is clear from both exercises, the use of single temperature models with fixed opacity and spectral emissivity parameters severely hampers the measurement of dust temperatures, by imposing strong biases and implicit uncertainties. We note that such a methodology is common in the literature, especially at high redshift \citep[e.g.][]{strandet_2016, faisst_2020, algera_2023_new_rebels}, where the dearth of FIR data necessitates simplifying assumptions and can bias any conclusions inferred from these models.

\section{Discussion}\label{sec:discussion}

 In this analysis, we have demonstrated challenges in measuring dust temperatures for a sample of $z\sim 6$ galaxies generated from a high resolution cosmological simulation. These galaxies are an ideal testing ground for dust temperature techniques, as they span a wide range of stellar, dust, and radiative properties at the earliest epochs. Below, we discuss our results in the context of previous dust temperature predictions from hydrodynamical simulations, dust detections of high-z galaxies, and the potential caveats owing to assumptions made when forward modeling our simulated galaxies. 

 \subsection{Comparisons to High-z Dust Temperatures from Hydrodynamical Models in the Literature}

The dust temperature modeling we present above demonstrates the array of challenges in characterizing the dust properties of high-z galaxies from both a forward modeling perspective and the backwards modeling perspective. For instance, while the Cosmic Sands galaxies benefit from having a self-consistent dust evolution model within {\sc simba}, we do need to make assumptions related to the optical properties of the dust grains within {\sc powderday}, the implications of which are described in further detail in \S \ref{sec:caveats}. And while we do not intend for this work to predict the dust temperatures of high-z galaxies, either in comparison to observational studies and other predictions from hydrodynamical models, but rather to highlight the difficulties in modeling dust properties, it would be useful to discuss the modeling choices made here in comparison to other simulations.

In recent years, several cosmological hydrodynamical galaxy formation models have studied the impact of dust on the evolution of galaxies, including galaxy chemical enrichment and dust mass evolution \citep[e.g.][]{mckinnon_2016, graziani_2020, choban_2022_fire, dicesare_2022, lewis_2022_dustier} and the effect of dust on the radiative properties of galaxies \citep[e.g.][]{liang_lichen_2019,mushtaq_2022, shen_2022_tng, vijayan_2022}. The former analysis necessitates the implementation of a dust evolution model within the galaxy formation framework (akin to {\sc simba}) while the latter can be studied using post-process radiative transfer models like {\sc skirt} \citep{skirt_1, skirt_2} and {\sc powderday} \citep{powderday}. Below we focus specifically on studies that presented dust temperature predictions for high-z galaxies.

In \cite{liang_lichen_2019}, the authors present dust temperature predictions from the {\sc MassiveFIRE} suite of simulations across a range of redshifts. The dust emission properties were modeled with {\sc skirt} with an adopted fiducial dust-to-metals (DTM) ratio of 0.4 and includes the impact of both diffuse dust and dust in stellar birth clouds, implemented with the {\sc mappingsIII} model. This is in contrast to Cosmic Sands in which we neglect sub-resolution radiative transfer processes. Focusing on the galaxies at $z=6$ (with stellar masses ranging from $10^{9} < $ M$_{\rm star} < 10^{11}$ M$_{\odot}$ and dust masses ranging from $10^{6.5} < $ M$_{\rm dust} < 10^{8}$ M$_{\odot}$), they predict mass-weighted dust temperatures ranging from $\sim 20 - 40$~K in contrast to the temperatures from Cosmic Sands which range from $30-80$~K with one galaxy at $\sim120$~K. Notably, however, the modified blackbody dust emission models the authors fit to mock photometry sampled from the {\sc skirt} SEDs struggled to reproduce the shape of the true SEDs, resulting in a large scatter in dust temperatures inferred from these fits, similar to the results for the Cosmic Sands galaxies shown in \S \ref{sec:measuring_dust_temps}. And though the predicted range of $z>6$ dust temperatures are different between the two simulations, they do agree that temperatures derived from modified blackbody SED fits tend to over-estimate the temperature corresponding to the peak of the FIR SED.
 
 While the IR luminosities and dust masses of the {\sc fire} galaxies are comparable to those of the Cosmic Sands galaxies, we note that the predicted (and directly modeled) global dust-to-metals ratio in the Cosmic Sands galaxies range from $\sim0.05-0.6$ with significant variations within each individual galaxy, in contrast to the value of $0.4$ \cite{liang_lichen_2019} assumed. Furthermore, the use of a the {\sc mappingsIII} model to account for dust emission in sub-resolution star forming regions can have significant impacts on the predicted dust temperatures, as we describe in \S \ref{sec:caveats}. Therefore, we conclude that differences in the post-processed treatment of dust emission and the treatment of the star-dust geometry/dust abundance drive the differences in the mass-weighted temperature distributions between the two models.

\cite{shen_2022_tng} presented high-z dust temperatures of massive ($10^{10} < $ M$_{\rm star} < 10^{11}$ M$_{\odot}$), IR luminous galaxies from the IllustrisTNG simulations. Dust absorption and emission are modeled with {\sc skirt}, assuming the redshift dependent dust-to-metal mass ratio presented in \cite{vogelsberger_2020} which is applied to cold, star-forming gas cells. Dust temperatures inferred from the peak of the FIR {\sc skirt} SEDs range between $\sim45-60$~K for galaxies at $z=6-8$, increasing to $\sim70-90$~K if measured from modified blackbody fits to the full FIR SED. While the authors describe both the peak temperature and the dust temperature inferred from modified blackbody fits as luminosity-weighted temperatures, it is important to note the caveats related to optical depths discussed in \S \ref{sec:optical_depth}. Compared to the Cosmic Sands galaxies, the TNG predicted dust temperatures derived from SED fits are comparable to the peak and luminosity-weighted temperatures we present in \S \ref{sec:mw_dust_temps}, perhaps owing to the more nuanced treatment of the dust abundance distribution in relation to star forming regions.

Finally, in \cite{vijayan_2022}, the authors present dust properties including temperatures for galaxies from the {\sc flares} simulation across a stellar mass range of $10^{8} < $ M$_{\rm star} < 10^{11}$ M$_{\odot}$. As above, dust absorption/emission is modeled with {\sc skirt}, with dust abundances scaled to the metal mass in cool ($T<10^6$~K) gas cells following the dust-to-metal mass relation of \cite{vijayan_2019}, which relates the gas phase ISM metallicity and the mass-weighted stellar population age to the dust abundance in each galaxy. For galaxies $z=6-10$, the dust-to-metals ratio ranges from $0.01-0.2$, which is much lower than both the assumed DTM of \cite{liang_lichen_2019} and the DTM values found in the Cosmic Sands galaxies but is closer to the redshift dependent DTM value used in \cite{shen_2022_tng} (DTM $\sim0.1$ at $z=6.5$). Also similar to \cite{shen_2022_tng}, the dust temperatures corresponding to the peak of the FIR SED are $\sim10-25$~K lower than those derived from modified blackbody fits to the entire SED. The latter dust temperatures are comparable to the luminosity-weighted temperatures of the Cosmic Sands galaxies and to those inferred from modified blackbody fits to photometry sampled from our {\sc powderday} SEDs.

\subsection{Implications for High-z Dust Temperature Measurements}

As we enter the joint ALMA-JWST era, evidence of greater-than-expected dust enrichment at the earliest epochs is continuing to push our understanding of galaxy formation \citep[e.g.][]{sommovigo_2022_rebels_alpine_dust, labbe_2023, akins_2023}.
The challenges in modeling dust temperatures for dusty, high-z galaxies outlined in this analysis serve to underscore the findings from previous studies spanning decades for both the Milky Way and nearby galaxies \citep{hildebrand_1983, dale_2002, draine_2006, casey_2012, kirkpatrick_2014, lee_2015} to galaxies at $z=2$ and beyond \citep{cortzen_2020, sommovigo_2022_rebels_alpine_dust, jin_2022} that sought to characterize the dust content in various systems despite the onslaught of systemic model degeneracies. These degeneracies, also explored by previous studies primarily focused on low redshift systems \citep[e.g.][]{klaas_2001, shetty_2009a, shetty_2009b, kelly_2012, rangwala_2011}, are amplified at high redshift, where the additional dearth of FIR data makes the characterization of dust problematic. The assumptions that we make when fitting FIR SEDs at low redshift may not hold for high redshift as shown in Section \S\ref{sec:measuring_dust_temps}, adding implicit uncertainties to already uncertain measurements.

Galaxies at these epochs that have relatively well sampled FIR SEDs have primarily been FIR- or submillimeter-selected dusty star-forming galaxies. For instance, the majority of $z>5$ South Pole Telescope (SPT) submillimeter galaxies (SMGs) have detections from \textit{Herschel}/SPIRE, APEX/LABOCA, SPT, and ALMA, effectively covering rest-frame $40-500\mu$m. In \cite{reuter_2020}, the authors constrain the dust properties for these galaxies by fitting a modified blackbody function to $4-6$ flux bands, depending on the galaxy, and account for optically thick emission in the FIR, similar to our Scenario 1. Like the estimates for the Cosmic Sands galaxies, the inferred blackbody dust temperatures for the SPT-SMG galaxies range between $30-70$~K, with a characteristic uncertainty of $\sim16$~K. Similar studies suggest the typical dust temperatures found in SMGs $\sim30-40$~K \citep[e.g.][]{greve_2012, swinbank_2014, strandet_2016, danielson_2017, sun_2022}. For UV-selected galaxies targeted by ALMA, studies tend to find higher dust temperatures ranging from $\sim40$ to $100$~K \citep[e.g.,][]{faisst_2020, bakx_2020, inami_2022, sommovigo_2022_rebels_alpine_dust, algera_2023_new_rebels}.

As shown above, for the relatively dust poor Cosmic Sands galaxies, the inferred temperatures generally align with the luminosity-weighted dust temperatures while for the dust rich galaxies, the inferred temperatures align with the mass-weighted. The lower apparent temperatures in high-z SMGs may also be the result of high dust opacities that obscure the bulk of the dust temperature distribution \citep{jin_2022}. We stress the importance of allowing general opacity solutions when inferring dust properties of high-z dusty galaxies. The relation between inferred and `true' dust temperatures depends on both the shape of the SED (which folds in the dust temperature distribution, the star-dust geometry, and the dust optical properties), and the aforementioned model assumptions. As noted in \S \ref{sec:mw_dust_temps}, there is typically not a 1:1 nor even linear relation between the apparent dust temperature and the mass-weighted or luminosity-weighted dust temperature of a galaxy. This is supported by the analyses of \cite{liang_lichen_2019} and \cite{vijayan_2022} from the {\sc fire} and {\sc flares} simulations, respectively, when relating apparent dust temperatures to the mass-weighted temperatures.

Inferring the dust properties of UV- or optically-selected galaxies at $z>6$ is potentially more challenging, as significant observation time must be dedicated to detecting faint dust emission. Large ALMA programs like ALPINE \citep{le_fevre_2020_alpine} and REBELs \citep{bouwens_2022_rebels} may only have a single dust continuum detection in the FIR (ALMA bands 6 or 7), which can lead to significant uncertainties when inferring dust temperatures as modified blackbody functions are not well constrained by the RJ tail. To constrain degenerate solutions, \cite{sommovigo_2021} proposed using [CII] detections to independently constrain the galaxy dust masses, thereby breaking the degeneracy between dust mass and temperature. In \cite{sommovigo_2022_rebels_temps, sommovigo_2022_rebels_alpine_dust}, the authors applied this methodology to high-z galaxies from the ALPINE and REBELs surveys, inferring dust temperatures by combining the rest-frame $158\mu$m [CII] and dust continuum detections, assuming a modified blackbody function and optically thin emission. The dust temperatures inferred from this method range from $30-60$~K with a median temperature for the REBELs galaxies of $47$~K \citep{sommovigo_2022_rebels_temps}. 

\cite{faisst_2020} fit the ALMA-detected FIR SEDs of four Keck/DEIMOS-selected $z=5-6$ galaxies in the COSMOS field, and are careful to point out that the dust temperatures measured from the SEDs are more similar to luminosity-weighted temperatures and may miss the bulk of dust mass in colder regions. Nevertheless, the galaxy SEDs are assumed to be optically thin past $100\mu$m, but constraining both the emissivity index and the dust temperature with rest-frame photometry at $\sim100-200\mu$m is challenging and significant uncertainties persist. The authors predict dust temperatures of $30-40$~K for these galaxies, matching the T$_{\rm d}$(z) relation predicted from hydrodynamical simulations presented in \cite{liang_lichen_2019}. These measurements also agree with our result from Scenario 2, in which we infer dust temperatures between $30-60$~K, albeit with relatively large uncertainties. Importantly, however, those temperatures do not correlate significantly with either the true mass-weighted or luminosity-weighted temperatures and so we caution the use of single-band dust continuum measurements to characterize the dust properties in observed galaxies.

\cite{fudamoto_2023} presented new estimates for high-z galaxies derived from ALMA observations, including A1689-zD1 \citep{watson_2015, knudsen_2017, bakx_2021}, B14-65666 \citep{hashimoto_2019, hashimoto_2022}, and MACS0416-Y1 \citep{tamura_2019, bakx_2020}. Their dust emission model, based on work by \cite{inoue_2020}, accounts for clumpiness in the ISM and enables the derivation of dust temperatures and luminosities from single-band dust continuum measurements. With this method, the authors found their temperature measurements generally agreed with existing literature estimates of dust temperatures \citep[e.g.,][]{sommovigo_2021, bakx_2020} but with relatively significant uncertainties from the dust/ISM geometry. 

\cite{witstok_2023} compile SEDs and measurements of dust emission areas of $17$ high redshift galaxies, including SMGs, ``normal" star-forming galaxies, and QSOs. They model the dust properties with modified blackbody fitting, assuming a general opacity model, by linking independently derived dust mass surface density measurements to the FIR emission. For galaxies at $z=6-8$, with stellar masses around $10^{10}$ M$_{\odot}$, they measure dust temperatures and IR luminosities similar to ULIRG systems at $z=0$, ranging from $35-80$~K, and find most of the galaxies in their sample are fit with an emissivity index of $\beta \sim 2$, similar to the dustier systems in the Cosmic Sands sample fit in Section \ref{sec:characterizing_sed}. The authors also discuss in detail the non-trivial nature of deriving dust properties of high-z galaxies, highlighting the complex relationship between measures of apparent temperature (T$_\mathrm{peak}$, T$_\mathrm{MBB}$) and physical temperatures (T$_\mathrm{mass-weighted}$, T$_\mathrm{luminosity-weighted}$).

Finally, \cite{algera_2023_new_rebels} presented new dust temperature measurements for 3 previously studied REBELs galaxies for which additional ALMA band 8 detections were obtained at rest-frame $\sim90\mu$m. With the leverage of a shorter wavelength detection, the authors fit modified blackbody functions with three techniques: $\beta=1.5$, $\beta=2.0$, and a variable $\beta$ for both optically thin and thick assumptions, in which the dust is assumed to transition from optically thick to thin at $\lambda_c = 100\mu$m ($\nu_c = 3000$~GHz). The temperatures inferred from these fits (assuming optically thin emission with $\beta=1.5$) are similar to the temperatures inferred in \cite{sommovigo_2022_rebels_temps} but with larger uncertainties. The temperatures inferred from the optically thick model are $10-15$~K larger than those from \cite{sommovigo_2022_rebels_temps}, implying that \textit{either} the assumption of optically thin emission at $\lambda > 40\mu$m is not valid for the REBELs galaxies \textit{and/or} the [CII]--derived dust mass constraint does not hold when an additional photometry constraints are available \citep[see, e.g., ][]{vizgan_2022}. 

What is clear is that the technique of inferring dust temperatures via modified blackbody fits to FIR SEDs is nontrivial for even well sampled SEDs, driven by degeneracies between dust properties and the fact that FIR emission may be optically thick out to several hundred micron, rendering a bulk of the dust mass invisible. The application of such models to high-z galaxies, where dust abundances and densities, star-dust geometries, and ISM conditions may be significantly different than low-z galaxies, while common, can introduce biases in measurements of dust properties as demonstrated in Section \S\ref{sec:measuring_dust_temps}. The relation between luminosity-weighted temperatures and mass-weighted temperatures depends on the star formation and dust properties of galaxies, making it challenging to apply a ``one-temperature-fits-all" model to a large sample of galaxies. The various techniques tested in this analysis demonstrate the large uncertainties on individual measurements, even when the FIR SED is relatively well constrained.

\subsection{Implications for High-z Dust Mass Measurements}

While our analysis has focused solely on the techniques used to infer galaxy dust temperatures from FIR continuum measurements, we briefly discuss the implications of these measurements, and their uncertainties, on their application to inferring dust masses. Complications arise when attempting to infer other dust properties from dust temperatures measured from the FIR SED; luminosity-weighted dust temperatures typically do not trace the mass-weighted dust temperature which is what is needed to estimate other dust properties such as mass. As pointed out by \cite{casey_2012} and \cite{cochrane_2022}, relatively small uncertainties on dust temperature result in significant errors on dust mass, especially when measured from photometry on the Rayleigh-Jeans tail that do little to constrain modified blackbody emission at shorter wavelengths. 

If we use the dust temperature estimates from Scenario 2 to infer dust masses, again assuming modified blackbody emission, the median offset is $-0.25$~dex -- nearly a factor of two -- from the true dust masses. The uncertainties propagated from the inferred dust temperature posteriors result in an average fractional error of $122\%$, significantly impacting our ability to conclusively determine the true dust masses of these galaxies. We highlight that this dust mass offset should be interpreted as a lower limit prediction for the actual offset between inferred and true dust masses for observed galaxies, as we are matching the dust optical properties ($\kappa, \beta$) between the backwards and forwards modeling of the dust emission. As explored in \cite{inoue_2020} and \cite{fanciullo_2020}, the adopted dust opacity and ISM geometry models have a significant impact on the derived dust masses even if the input dust temperature is correct, leading to inconsistencies in mass measurements between methodologies \citep{jin_2022}.

\begin{figure}
    \centering
    \includegraphics[width=0.47\textwidth]{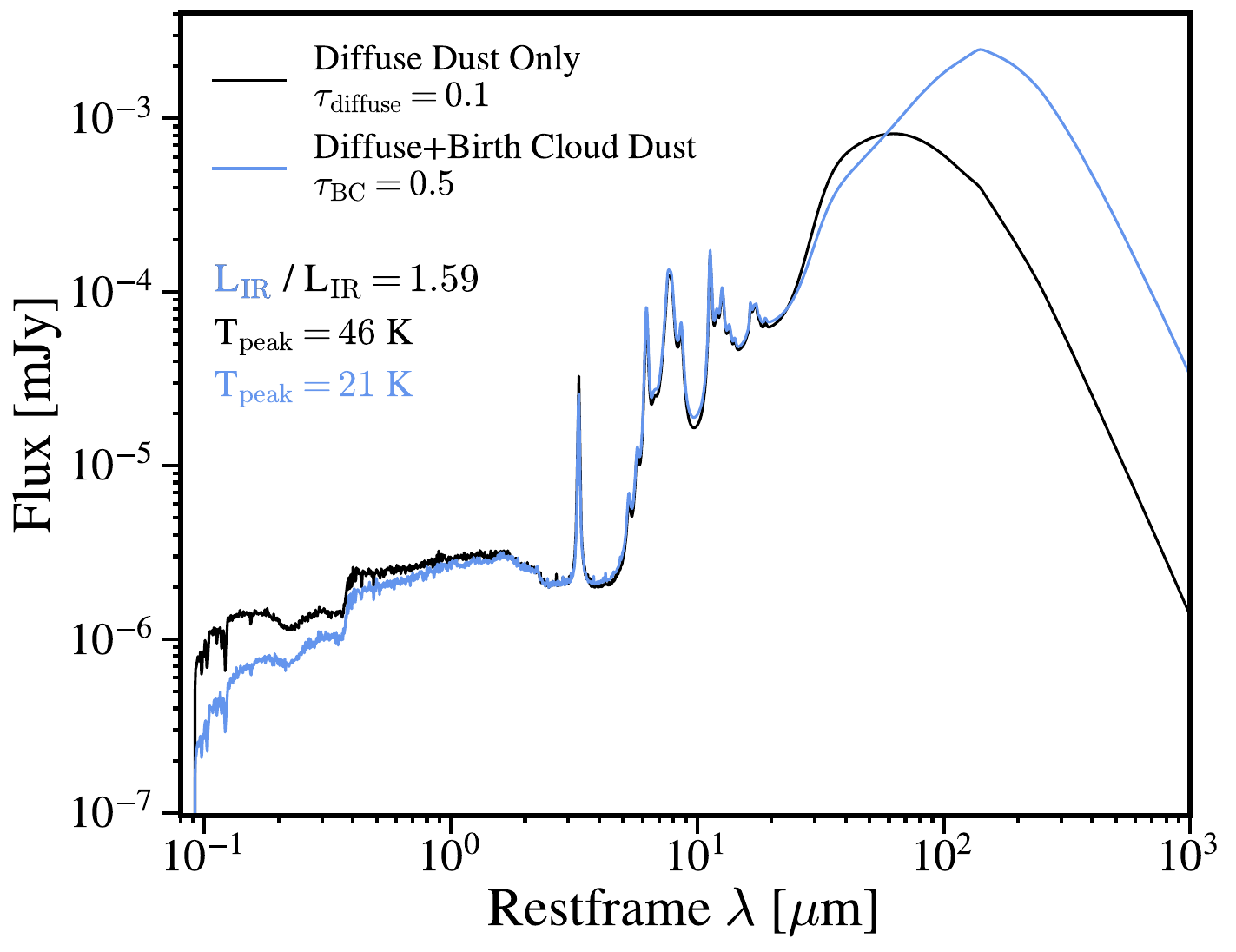}\\
    \includegraphics[width=0.47\textwidth]{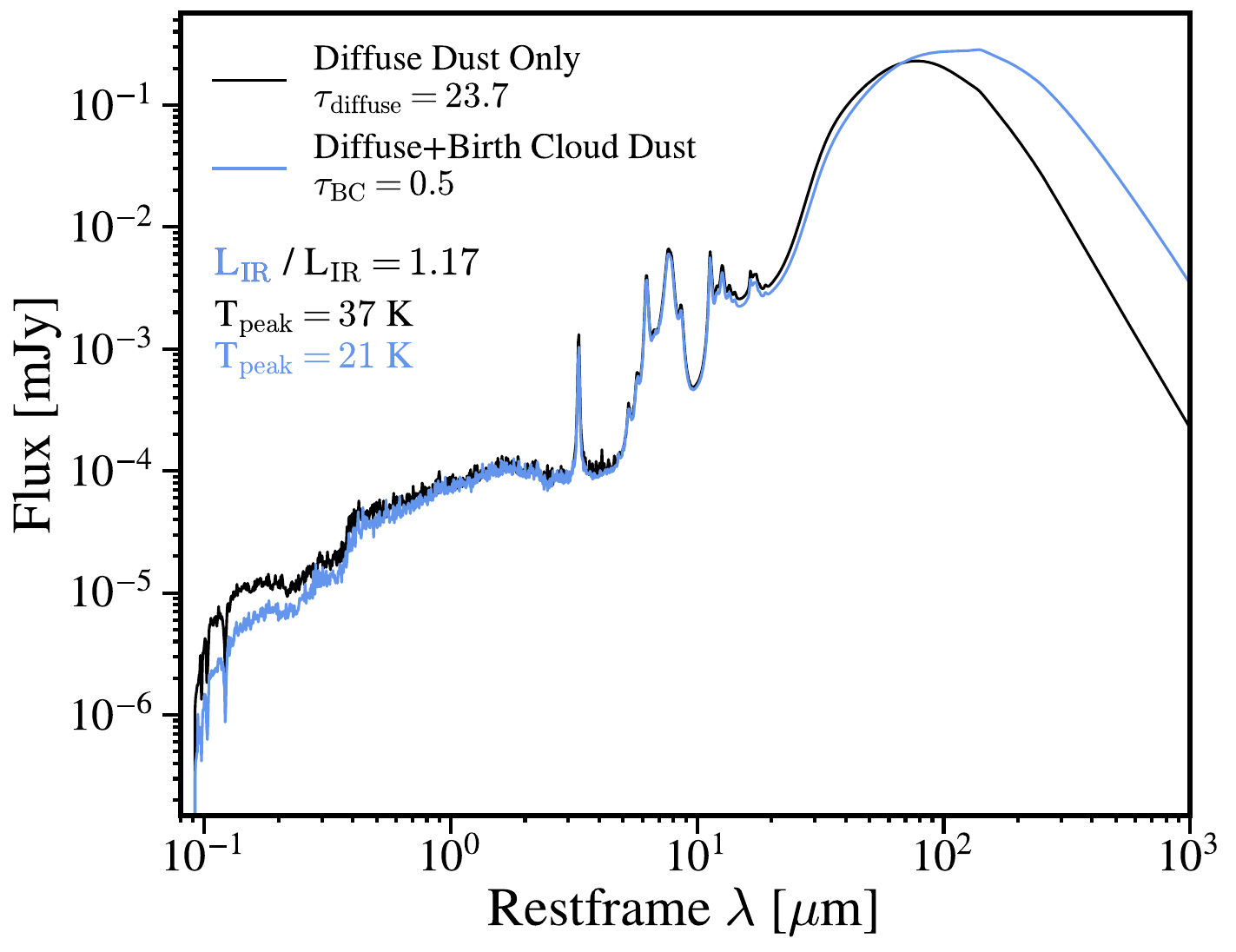}
    \caption{SEDs with (blue curves) and without (black curves) sub-resolution dust absorption and emission for two Cosmic Sands galaxies chosen to roughly span the range of dust optical depths at $100\mu$m. The galaxy in the top panel has a relatively low dust optical depth and the addition of a birth cloud model significantly changes the shape of the FIR SED, increases the IR luminosity by a factor of 1.6, and decreases the peak temperature by $25$~K. The bottom panel shows a galaxy with higher dust optical depth and the changes to the IR luminosity and peak temperature by adding birth cloud dust are less severe but still meaningful.}
    \label{fig:birth_cloud}
\end{figure}

\subsection{Caveats to Our Model}\label{sec:caveats}

A major limiting factor in the predictive power of the Cosmic Sands galaxies is the reliance on sub resolution and post process modeling of dust emission properties, including the treatment of dust in stellar birth clouds and the lack of on-the-fly ionized radiative transfer in HII and photo-dissociation regions. In this work, we generated UV-FIR SEDs with post-process radiative transfer but neglect additional attenuation terms towards young stars in birth clouds; i.e., we model only the diffuse dust attenuation and emission. The inclusion of a sub-resolution dust model \citep[e.g.,][in which young stars are attenuated by an additional birth cloud dust component]{charlot_fall_2000}, can significantly change the shape of the FIR SED as well as the dust temperature distribution. 

For instance, placing a dust screen with $\tau_V=0.5$~mag in front of stars $<10$~Myr old increases the total IR luminosity by a factor of $1.3-2$ on average across the Cosmic Sands galaxies and can shift the dust temperature corresponding to the peak of the FIR SED down by nearly $30$~K as less UV light is reaching the diffuse dust. We highlight two galaxies in Figure \ref{fig:birth_cloud} to demonstrate this scenario. The two galaxies have relatively low (top panel) and high (bottom panel) dust optical depths at $100\mu$m. The black curves show the fiducial SEDs and the blue curves show the impact of adding a uniform dust screen in front of young stars. For the low dust opacity galaxy, the addition of birth clouds significantly changes the shape of the FIR SED, increases the IR luminosity by a factor of 1.6, and decreases the peak temperature by $25$~K. On the other hand, the high dust opacity galaxy experiences a modest increase in IR luminosity with the peak of the FIR SED shifting slightly to higher wavelengths. In this regard, the dust temperatures presented here may not be directly comparable to dust temperature distributions of real galaxies, or to the trends found in other works analyzing the dust temperatures of high-z galaxies from hydrodynamical simulations \citep[e.g.][]{liang_lichen_2019}. However, these model assumptions will not significantly impact the comparisons between the characteristic dust temperatures (luminosity- and mass-weighted) and the temperatures inferred from modified blackbody fits in our analysis as both are related to the same post-process radiative transfer assumptions. 

A perhaps more fundamental uncertainty is the underlying galaxy formation physics in {\sc simba}, namely the treatment of stellar sources of feedback on the ISM. A significant obstacle in numerically modeling galaxy evolution is the limitation of resolution such that star-formation will not be self-regulating at the smallest scales. To get around this, {\sc simba} adopts an artificial equation of state (EoS) that sets a pressure floor which suppresses fragmentation below the scale of the smoothing volume \citep{dave_2016, dave_simba}. This is in contrast to models such as {\sc FIRE} \citep{hopkins_FIRE_og} and {\sc smuggle} \citep{marinacci_2019_smuggle}, in which star formation and feedback from stellar sources is explicitly resolved, allowing the multi-phase medium to be more accurately modeled.

In the context of this analysis, the nonlinear coupling of stellar feedback modes on the surrounding medium can play a significant role in the predicted dust evolution, star-dust geometry, and dust temperature distribution of high-z galaxies \citep{esmerian_2023}. For instance, in the {\sc simba} model, the mass-loading factor of the star formation-driven galactic winds is artificially reduced at $z>3$ to better match observations of galaxy growth at high redshift \citep{dave_simba}. This could lead to the pile up of gas in the nuclear regions of high-z galaxies, yielding large surface densities and dust abundances that could otherwise be dispelled by stellar winds. The consequences of these fundamental, theoretical uncertainties clearly add to the overall uncertainty budget when predicting dust temperatures from hydrodynamical models.

Lastly, we are unable to predict the impact of variable dust extinction on the FIR emission of high-z galaxies. Specifically, we do not model an evolving grain-size distribution in {\sc simba}, which could significantly change the thermodynamics of the ISM in these galaxies and drive changes in the ways dust absorbs and emits as a function of wavelength. It is known that dust of differing grain sizes have different optical properties and thus dominate galaxy spectra at different wavelengths; e.g., polycyclic aromatic hydrocarbons with grain sizes $\sim 1-10$\textup{~\AA} dominate galaxy emission in the near- to mid-IR \citep[e.g.][]{narayanan_2023}. While we cannot test the impact of galaxy dust grain size distributions and how this manifests in IR SEDs, we predict that the addition of a self-consistent grain size evolution model, resulting in non-universal dust optical properties and perhaps a wider diversity in FIR SED shapes, would result in greater uncertainties in the inferred dust temperatures from modified blackbody fits to FIR SEDs. 

\subsection{Outlook and Paths Forward}

The question remains of how can we go forward from this analysis, given the non trivial nature of measuring dust temperature and the limited relation this temperature has with the real, physical distribution of dust temperatures within a galaxy. If one's goal was simply to estimate some representative dust temperature, the results from Section \ref{sec:fitting_fir_phot} demonstrate the necessity of robust photometric coverage of the FIR SED. Even then, the utility of this temperature measurement is unclear; the relation between it and the physical temperatures describing the underlying distribution are rarely 1:1, rendering any measure of, e.g., dust mass or the thermal state of the ISM, highly uncertain.


If instead the goal of measuring dust properties is to gain a better understanding of the galaxy as a whole, it would be more powerful to place FIR continuum observations in the context of the entire SED, including nebular and molecular emission lines where available, to jointly map UV-FIR photometry to the underlying physical properties. The constraints on dust and ISM properties from UV and optical photometry (e.g., dust composition, grain size distribution, metallicity, star-dust geometry) are complemented by the constraints in the FIR (gas mass, dust mass and temperature distribution), allowing one to estimate galaxy physical properties simultaneously and with respect to the correlations that physically exist between galaxy components (e.g., the dust emission in the FIR will be influenced by the model SFH and UV-optical attenuation). Balancing the energy transfer between starlight and dust provides more information than either measurement gives individually.

In other words, the exercises conducted here have limited utility in terms of offering solutions for improving dust temperature measurements. This said, studies like \cite{haskell_2023}, \cite{jones_2022} and \cite{jones_2023} highlight the benefits of a panchromatic perspective, enforcing symmetry between dust properties modeled in both the UV-optical regime and in the FIR. \cite{jones_2022} fit IR-luminous galaxies from COSMOS ($0 < z < 0.5$), with \cite{jones_2023} extending to higher redshifts ($z < 6$), and find that the inferred dust and stellar properties are mutually dependent on the assumed SFH, stellar population synthesis model, and dust emission model. In testing the utility of the energy balance constraint on high-z galaxies simulated with {\sc fire}, \cite{haskell_2023} demonstrate improvements in inferring dust mass and IR luminosity from panchromatic photometry compared to estimates from FIR photometry alone. 

Of course, there are significant caveats to this methodology. For instance, as the model parameter space increases, unless robustly constrained by high fidelity spectro-photometric observations, uncertainties will inflate to reflect the degree to which degenerate solutions exist. Moreover, the current survey of high-z galaxies is dominated by galaxies selected based on rest-frame UV-optical or FIR emission thus many systems are missing complementary photometry in other bands. In this case, relying on bright emission lines in the FIR, like CO or [CII] can be beneficial when trying to characterize the thermodynamics of the ISM \citep[e.g.,][]{reuter_2023}. And though many studies have been devoted to robustly modeling stellar emission with sophisticated implementations used in codes like {\sc prospector} \citep{johnson_2021} and {\sc bagpipes} \citep{carnall_2018}, the same is only marginally true for dust emission, which is primarily based on course-grid templates like \cite{draine_infrared_2007} and potentially ineffective for high-z galaxies. 

Lastly, hydrodynamical simulations can be used to predict the mapping between physical properties and observables. For instance, \cite{cochrane_2022} modeled the FIR emission from galaxies from the {\sc fire} simulation suite and found tight relations between certain flux density ratios and the mass-weighted temperatures of both star-forming and quenched galaxies, though this is highly dependent on the underlying physics of the {\sc fire} galaxy formation model and the details of the forward modeling of the dust emission. Similarly, etechniques instead using machine learning algorithms trained on simulations to find the nonlinear mappings between luminosity and physical properties also show promise: indeed \cite{gilda_2021} demonstrate with {\sc mirkwood} significant improvements in inferring galaxy physical properties from broadband photometry compared to traditional parametric SED modeling across a range of galaxy types and masses. This said, similar to the \citet{cochrane_2022} results, the \citet{gilda_2021} machine learning model is  dependent on the simulations and forward modeling methods used to train the machine learning algorithm.

\section{Conclusion}

Using galaxies from the Cosmic Sands suite of massive, dusty galaxies in the EoR, we have demonstrated the challenges in modeling dust temperatures at high redshift, both forwards from hydrodynamical simulations post-processed with 3D dust radiative transfer and backwards from modified blackbody fits to broadband FIR photometry. We first examined the ``true" dust temperatures of the Cosmic Sands galaxies, calculated from {\sc powderday}'s radiative transfer coupled with the {\sc simba} dust evolution model. Intuitively, the dust temperature distributions are diverse from galaxy-to-galaxy, but also within each galaxy as shown in Figure \ref{fig:temp_dist}, with mass- and luminosity-weighted temperatures producing unique distributions for each galaxy. As shown in Figures \ref{fig:Tmw_Tlw} and \ref{fig:temp_lir_tau}, the star-dust geometry and ISM column densities contribute to this diversity by breaking correlations between intrinsic and apparent dust properties for dustier, star-forming galaxies. It also makes comparisons between theoretical predictions and observations of dust properties challenging as the mapping between intrinsic temperature and observable temperature depends on each galaxy's properties like mass, star-dust geometry, and age. Lastly, in an attempt to characterize the full dust FIR SEDs with a modified blackbody model, we find that the Cosmic Sands FIR dust opacities are not well described by the commonly assumed powerlaw relation with emission frequency, $\tau = (\nu / \nu_0)^{\beta}$.

Then, starting from the {\sc powderday} FIR SEDs, we attempted to infer dust temperatures as an observer would: we ran two cases that differed by SED coverage, spanning the range of photometry available for $z=7$ galaxies, and by model assumptions regarding the dust optical depth. Our results, summarized in Figures \ref{fig:Tbb_full} and \ref{fig:Tbb_2}, demonstrate the difficulties in constraining dust temperatures. The inherent degeneracies between dust properties (temperature, mass, star-dust geometry, dust grain size, composition, and structure) make it challenging to robustly constrain the dust temperature even for well sampled FIR SEDs \citep{kovacs_2010, casey_2012}. As is clear from this exercise, fitting FIR SEDs with a single temperature component and a simple parameterization for the dust optical depth does not work for a majority of high-z galaxies. Until high-z galaxy FIR SEDs can be more robustly detected over a wide wavelength range, these uncertainties pertaining to dust properties will persist.

From this analysis, we have shown that both theoretical and observational characterizations of galaxy dust properties are fraught with uncertainties at high-z. We note that predictions for dust temperatures at high redshift are highly dependent on model assumptions regarding dust evolution (e.g., dust abundances and star-dust geometry) and dust radiative transfer (e.g., sub-resolution dust in star-forming regions as shown in in Figure \ref{fig:birth_cloud}, dust optical properties) and many of these factors remain relatively unconstrained or difficult to implement in a self-consistent manner. However, results from observational studies of high-z galaxies can be highly uncertain, both implicitly and explicitly, depending on the model assumptions and the ability to constrain FIR dust emission with sparse photometric data. Thus, any prediction for high-z dust properties from theoretical models should be placed in the appropriate context with regard to specific model assumptions and caution should be taken when comparing theoretical predictions to results from observational studies.

\acknowledgements
The Cosmic Dawn Center is funded by the Danish National Research Foundation under grant No. 140. SL and DN acknowledge support from the NSF via grant AST-1909153. This work was initiated at the Aspen Center for Physics, which is supported by NSF grant PHY-1607611. The National Radio Astronomy Observatory is a facility of the National Science Foundation operated under cooperative agreement by Associated Universities, Inc.

The {\sc simba} simulations used in this work were run on the Hipergator computing cluster at the University of Florida. The flagship {\sc simba} simulations were run on the ARCHER U.K. National Supercomputing Service {\tt http://www.archer.ac.uk}.

\textit{Software}: {\sc simba} \citep{dave_simba}, {\sc music} \citep{hahn_abel_2011_music}, {\sc caesar} \citep{caesar}, {\sc powderday} (\citealt{powderday}), {\sc hyperion} \citep{robitaille_2011_hyperion}, {\sc MCIRSED} \citep{drew_2022}, {\sc PyMC3} \citep{salvatier_2016}, {\sc emcee} \citep{emcee}, {\sc numpy} \citep{harris2020array_numpy}, {\sc matplotlib} \citep{Hunter_2007, 2018zndo...1482098C}, {\sc pandas} \citep{mckinney-proc-scipy-2010, reback2020pandas}, {\sc yt} \citep{turk_2011_yt}, {\sc sphviewer} \citep{alejandro_benitez_llambay_2015_21703}

\bibliography{bib}{}

\begin{thebibliography}{}
\expandafter\ifx\csname natexlab\endcsname\relax\def\natexlab#1{#1}\fi
\providecommand{\url}[1]{\href{#1}{#1}}
\providecommand{\dodoi}[1]{doi:~\href{http://doi.org/#1}{\nolinkurl{#1}}}
\providecommand{\doeprint}[1]{\href{http://ascl.net/#1}{\nolinkurl{http://ascl.net/#1}}}
\providecommand{\doarXiv}[1]{\href{https://arxiv.org/abs/#1}{\nolinkurl{https://arxiv.org/abs/#1}}}

\bibitem[{{Aalto} {et~al.}(2019){Aalto}, {Muller}, {K{\"o}nig}, {Falstad},
  {Mangum}, {Sakamoto}, {Privon}, {Gallagher}, {Combes}, {Garc{\'\i}a-Burillo},
  {Mart{\'\i}n}, {Viti}, {van der Werf}, {Evans}, {Black}, {Varenius},
  {Beswick}, {Fuller}, {Henkel}, {Kohno}, {Alatalo}, \&
  {M{\"u}hle}}]{aalto_2019}
{Aalto}, S., {Muller}, S., {K{\"o}nig}, S., {et~al.} 2019, \aap, 627, A147,
  \dodoi{10.1051/0004-6361/201935480}

\bibitem[{{Akins} {et~al.}(2023){Akins}, {Casey}, {Allen}, {Bagley},
  {Dickinson}, {Finkelstein}, {Franco}, {Harish}, {Arrabal Haro}, {Ilbert},
  {Kartaltepe}, {Koekemoer}, {Liu}, {Long}, {McCracken}, {Paquereau},
  {Papovich}, {Pirzkal}, {Rhodes}, {Robertson}, {Shuntov}, {Toft}, {Yang},
  {Barro}, {Bisigello}, {Buat}, {Champagne}, {Cooper}, {Costantin}, {de la
  Vega}, {Drakos}, {Faisst}, {Fontana}, {Fujimoto}, {Gillman},
  {G{\'o}mez-Guijarro}, {Gozaliasl}, {Hathi}, {Hayward}, {Hirschmann},
  {Holwerda}, {Jin}, {Kocevski}, {Kokorev}, {Lambrides}, {Lucas}, {Magdis},
  {Magnelli}, {McKinney}, {Mobasher}, {P{\'e}rez-Gonz{\'a}lez}, {Rich},
  {Seill{\'e}}, {Talia}, {Urry}, {Valentino}, {Whitaker}, {Yung}, \&
  {Zavala}}]{akins_2023}
{Akins}, H.~B., {Casey}, C.~M., {Allen}, N., {et~al.} 2023, arXiv e-prints,
  arXiv:2304.12347, \dodoi{10.48550/arXiv.2304.12347}

\bibitem[{{Algera} {et~al.}(2023){Algera}, {Inami}, {Sommovigo}, {Fudamoto},
  {Schneider}, {Graziani}, {Dayal}, {Bouwens}, {Aravena}, {da Cunha},
  {Ferrara}, {Hygate}, {van Leeuwen}, {De Looze}, {Palla}, {Pallottini},
  {Smit}, {Stefanon}, {Topping}, \& {van der Werf}}]{algera_2023_new_rebels}
{Algera}, H., {Inami}, H., {Sommovigo}, L., {et~al.} 2023, arXiv e-prints,
  arXiv:2301.09659, \dodoi{10.48550/arXiv.2301.09659}

\bibitem[{{Angl{\'e}s-Alc{\'a}zar} {et~al.}(2017){Angl{\'e}s-Alc{\'a}zar},
  {Faucher-Gigu{\`e}re}, {Quataert}, {Hopkins}, {Feldmann}, {Torrey}, {Wetzel},
  \& {Kere{\v{s}}}}]{angles-alcazar_FIRE}
{Angl{\'e}s-Alc{\'a}zar}, D., {Faucher-Gigu{\`e}re}, C.-A., {Quataert}, E.,
  {et~al.} 2017, \mnras, 472, L109, \dodoi{10.1093/mnrasl/slx161}

\bibitem[{{Bakx} {et~al.}(2020){Bakx}, {Tamura}, {Hashimoto}, {Inoue}, {Lee},
  {Mawatari}, {Ota}, {Umehata}, {Zackrisson}, {Hatsukade}, {Kohno}, {Matsuda},
  {Matsuo}, {Okamoto}, {Shibuya}, {Shimizu}, {Taniguchi}, \&
  {Yoshida}}]{bakx_2020}
{Bakx}, T. J.~L.~C., {Tamura}, Y., {Hashimoto}, T., {et~al.} 2020, \mnras, 493,
  4294, \dodoi{10.1093/mnras/staa509}

\bibitem[{{Bakx} {et~al.}(2021){Bakx}, {Sommovigo}, {Carniani}, {Ferrara},
  {Akins}, {Fujimoto}, {Hagimoto}, {Knudsen}, {Pallottini}, {Tamura}, \&
  {Watson}}]{bakx_2021}
{Bakx}, T. J.~L.~C., {Sommovigo}, L., {Carniani}, S., {et~al.} 2021, \mnras,
  508, L58, \dodoi{10.1093/mnrasl/slab104}

\bibitem[{Benitez-Llambay(2015)}]{alejandro_benitez_llambay_2015_21703}
Benitez-Llambay, A. 2015, py-sphviewer: Py-SPHViewer v1.0.0,
  \dodoi{10.5281/zenodo.21703}

\bibitem[{{Blain} {et~al.}(2003){Blain}, {Barnard}, \& {Chapman}}]{blain_2003}
{Blain}, A.~W., {Barnard}, V.~E., \& {Chapman}, S.~C. 2003, \mnras, 338, 733,
  \dodoi{10.1046/j.1365-8711.2003.06086.x}

\bibitem[{{Bouwens} {et~al.}(2022){Bouwens}, {Smit}, {Schouws}, {Stefanon},
  {Bowler}, {Endsley}, {Gonzalez}, {Inami}, {Stark}, {Oesch}, {Hodge},
  {Aravena}, {da Cunha}, {Dayal}, {de Looze}, {Ferrara}, {Fudamoto},
  {Graziani}, {Li}, {Nanayakkara}, {Pallottini}, {Schneider}, {Sommovigo},
  {Topping}, {van der Werf}, {Algera}, {Barrufet}, {Hygate}, {Labb{\'e}},
  {Riechers}, \& {Witstok}}]{bouwens_2022_rebels}
{Bouwens}, R.~J., {Smit}, R., {Schouws}, S., {et~al.} 2022, \apj, 931, 160,
  \dodoi{10.3847/1538-4357/ac5a4a}

\bibitem[{{Camps} \& {Baes}(2015)}]{skirt_1}
{Camps}, P., \& {Baes}, M. 2015, Astronomy and Computing, 9, 20,
  \dodoi{10.1016/j.ascom.2014.10.004}

\bibitem[{{Camps} \& {Baes}(2020)}]{skirt_2}
---. 2020, Astronomy and Computing, 31, 100381,
  \dodoi{10.1016/j.ascom.2020.100381}

\bibitem[{{Carnall} {et~al.}(2018){Carnall}, {McLure}, {Dunlop}, \&
  {Dav{\'e}}}]{carnall_2018}
{Carnall}, A.~C., {McLure}, R.~J., {Dunlop}, J.~S., \& {Dav{\'e}}, R. 2018,
  \mnras, 480, 4379, \dodoi{10.1093/mnras/sty2169}

\bibitem[{{Casey}(2012)}]{casey_2012}
{Casey}, C.~M. 2012, \mnras, 425, 3094,
  \dodoi{10.1111/j.1365-2966.2012.21455.x}

\bibitem[{{Caswell} {et~al.}(2018){Caswell}, {Droettboom}, {Hunter}, {Firing},
  {Lee}, {Stansby}, {Sales de Andrade}, {Hedegaard Nielsen}, {Klymak},
  {Varoquaux}, {Root}, {Elson}, {Dale}, {May}, {Lee}, {Sepp{\"a}nen},
  {Hoffmann}, {McDougall}, {Straw}, {Hobson}, {cgohlke}, {Yu}, {Ma}, {Vincent},
  {Silvester}, {Moad}, {Katins}, {Kniazev}, {Ariza}, \&
  {W{\"u}rtz}}]{2018zndo...1482098C}
{Caswell}, T.~A., {Droettboom}, M., {Hunter}, J., {et~al.} 2018,
  {Matplotlib/Matplotlib V3.0.1}, v3.0.1,  Zenodo,
  \dodoi{10.5281/zenodo.1482098}

\bibitem[{{Charlot} \& {Fall}(2000)}]{charlot_fall_2000}
{Charlot}, S., \& {Fall}, S.~M. 2000, \apj, 539, 718, \dodoi{10.1086/309250}

\bibitem[{{Choban} {et~al.}(2022){Choban}, {Kere{\v{s}}}, {Hopkins},
  {Sandstrom}, {Hayward}, \& {Faucher-Gigu{\`e}re}}]{choban_2022_fire}
{Choban}, C.~R., {Kere{\v{s}}}, D., {Hopkins}, P.~F., {et~al.} 2022, \mnras,
  514, 4506, \dodoi{10.1093/mnras/stac1542}

\bibitem[{{Choi} {et~al.}(2016){Choi}, {Dotter}, {Conroy}, {Cantiello},
  {Paxton}, \& {Johnson}}]{mist_1}
{Choi}, J., {Dotter}, A., {Conroy}, C., {et~al.} 2016, \apj, 823, 102,
  \dodoi{10.3847/0004-637X/823/2/102}

\bibitem[{{Clements} {et~al.}(2010){Clements}, {Dunne}, \&
  {Eales}}]{clements_2010}
{Clements}, D.~L., {Dunne}, L., \& {Eales}, S. 2010, \mnras, 403, 274,
  \dodoi{10.1111/j.1365-2966.2009.16064.x}

\bibitem[{{Cochrane} {et~al.}(2022){Cochrane}, {Hayward}, \&
  {Angl{\'e}s-Alc{\'a}zar}}]{cochrane_2022}
{Cochrane}, R.~K., {Hayward}, C.~C., \& {Angl{\'e}s-Alc{\'a}zar}, D. 2022,
  \apjl, 939, L27, \dodoi{10.3847/2041-8213/ac951d}

\bibitem[{{Conroy} \& {Gunn}(2010)}]{fsps_2}
{Conroy}, C., \& {Gunn}, J.~E. 2010, \apj, 712, 833,
  \dodoi{10.1088/0004-637X/712/2/833}

\bibitem[{{Conroy} {et~al.}(2009){Conroy}, {Gunn}, \& {White}}]{fsps_1}
{Conroy}, C., {Gunn}, J.~E., \& {White}, M. 2009, \apj, 699, 486,
  \dodoi{10.1088/0004-637X/699/1/486}

\bibitem[{{Cortzen} {et~al.}(2020){Cortzen}, {Magdis}, {Valentino}, {Daddi},
  {Liu}, {Rigopoulou}, {Sargent}, {Riechers}, {Cormier}, {Hodge}, {Walter},
  {Elbaz}, {B{\'e}thermin}, {Greve}, {Kokorev}, \& {Toft}}]{cortzen_2020}
{Cortzen}, I., {Magdis}, G.~E., {Valentino}, F., {et~al.} 2020, \aap, 634, L14,
  \dodoi{10.1051/0004-6361/201937217}

\bibitem[{{Dale} \& {Helou}(2002)}]{dale_2002}
{Dale}, D.~A., \& {Helou}, G. 2002, \apj, 576, 159, \dodoi{10.1086/341632}

\bibitem[{{Danielson} {et~al.}(2011){Danielson}, {Swinbank}, {Smail}, {Cox},
  {Edge}, {Weiss}, {Harris}, {Baker}, {De Breuck}, {Geach}, {Ivison}, {Krips},
  {Lundgren}, {Longmore}, {Neri}, \& {Flaquer}}]{danielson_2011}
{Danielson}, A.~L.~R., {Swinbank}, A.~M., {Smail}, I., {et~al.} 2011, \mnras,
  410, 1687, \dodoi{10.1111/j.1365-2966.2010.17549.x}

\bibitem[{{Danielson} {et~al.}(2017){Danielson}, {Swinbank}, {Smail},
  {Simpson}, {Casey}, {Chapman}, {da Cunha}, {Hodge}, {Walter}, {Wardlow},
  {Alexander}, {Brandt}, {de Breuck}, {Coppin}, {Dannerbauer}, {Dickinson},
  {Edge}, {Gawiser}, {Ivison}, {Karim}, {Kovacs}, {Lutz}, {Menten},
  {Schinnerer}, {Wei{\ss}}, \& {van der Werf}}]{danielson_2017}
---. 2017, \apj, 840, 78, \dodoi{10.3847/1538-4357/aa6caf}

\bibitem[{{Dav{\'e}} {et~al.}(2019){Dav{\'e}}, {Angl{\'e}s-Alc{\'a}zar},
  {Narayanan}, {Li}, {Rafieferantsoa}, \& {Appleby}}]{dave_simba}
{Dav{\'e}}, R., {Angl{\'e}s-Alc{\'a}zar}, D., {Narayanan}, D., {et~al.} 2019,
  \mnras, 486, 2827, \dodoi{10.1093/mnras/stz937}

\bibitem[{{Dav{\'e}} {et~al.}(2016){Dav{\'e}}, {Thompson}, \&
  {Hopkins}}]{dave_2016}
{Dav{\'e}}, R., {Thompson}, R., \& {Hopkins}, P.~F. 2016, \mnras, 462, 3265,
  \dodoi{10.1093/mnras/stw1862}

\bibitem[{{Di Cesare} {et~al.}(2022){Di Cesare}, {Graziani}, {Schneider},
  {Ginolfi}, {Venditti}, {Santini}, \& {Hunt}}]{dicesare_2022}
{Di Cesare}, C., {Graziani}, L., {Schneider}, R., {et~al.} 2022, arXiv
  e-prints, arXiv:2209.05496.
\newblock \doarXiv{2209.05496}

\bibitem[{{Dotter}(2016)}]{mist_2}
{Dotter}, A. 2016, \apjs, 222, 8, \dodoi{10.3847/0067-0049/222/1/8}

\bibitem[{{Draine}(2003)}]{draine_03_araa}
{Draine}, B.~T. 2003, \araa, 41, 241,
  \dodoi{10.1146/annurev.astro.41.011802.094840}

\bibitem[{{Draine}(2006)}]{draine_2006}
---. 2006, \apj, 636, 1114, \dodoi{10.1086/498130}

\bibitem[{{Draine} \& {Lee}(1984)}]{draine_lee_1984}
{Draine}, B.~T., \& {Lee}, H.~M. 1984, \apj, 285, 89, \dodoi{10.1086/162480}

\bibitem[{{Draine} \& {Li}(2007)}]{draine_infrared_2007}
{Draine}, B.~T., \& {Li}, A. 2007, \apj, 657, 810, \dodoi{10.1086/511055}

\bibitem[{{Drew} \& {Casey}(2022)}]{drew_2022}
{Drew}, P.~M., \& {Casey}, C.~M. 2022, \apj, 930, 142,
  \dodoi{10.3847/1538-4357/ac6270}

\bibitem[{{Dunne} \& {Eales}(2001)}]{dunne_2001}
{Dunne}, L., \& {Eales}, S.~A. 2001, \mnras, 327, 697,
  \dodoi{10.1046/j.1365-8711.2001.04789.x}

\bibitem[{{Esmerian} \& {Gnedin}(2023)}]{esmerian_2023}
{Esmerian}, C.~J., \& {Gnedin}, N.~Y. 2023, arXiv e-prints, arXiv:2308.11723.
\newblock \doarXiv{2308.11723}

\bibitem[{{Faisst} {et~al.}(2020){Faisst}, {Fudamoto}, {Oesch}, {Scoville},
  {Riechers}, {Pavesi}, \& {Capak}}]{faisst_2020}
{Faisst}, A.~L., {Fudamoto}, Y., {Oesch}, P.~A., {et~al.} 2020, \mnras, 498,
  4192, \dodoi{10.1093/mnras/staa2545}

\bibitem[{{Falstad} {et~al.}(2021){Falstad}, {Aalto}, {K{\"o}nig}, {Onishi},
  {Muller}, {Gorski}, {Sato}, {Stanley}, {Combes}, {Gonz{\'a}lez-Alfonso},
  {Mangum}, {Evans}, {Barcos-Mu{\~n}oz}, {Privon}, {Linden},
  {D{\'\i}az-Santos}, {Mart{\'\i}n}, {Sakamoto}, {Harada}, {Fuller},
  {Gallagher}, {van der Werf}, {Viti}, {Greve}, {Garc{\'\i}a-Burillo},
  {Henkel}, {Imanishi}, {Izumi}, {Nishimura}, {Ricci}, \&
  {M{\"u}hle}}]{falstad_2021}
{Falstad}, N., {Aalto}, S., {K{\"o}nig}, S., {et~al.} 2021, \aap, 649, A105,
  \dodoi{10.1051/0004-6361/202039291}

\bibitem[{{Fanciullo} {et~al.}(2020){Fanciullo}, {Kemper}, {Scicluna},
  {Dharmawardena}, \& {Srinivasan}}]{fanciullo_2020}
{Fanciullo}, L., {Kemper}, F., {Scicluna}, P., {Dharmawardena}, T.~E., \&
  {Srinivasan}, S. 2020, \mnras, 499, 4666, \dodoi{10.1093/mnras/staa2911}

\bibitem[{{Foreman-Mackey} {et~al.}(2013){Foreman-Mackey}, {Hogg}, {Lang}, \&
  {Goodman}}]{emcee}
{Foreman-Mackey}, D., {Hogg}, D.~W., {Lang}, D., \& {Goodman}, J. 2013, PASP,
  125, 306, \dodoi{10.1086/670067}

\bibitem[{{Fudamoto} {et~al.}(2023){Fudamoto}, {Inoue}, \&
  {Sugahara}}]{fudamoto_2023}
{Fudamoto}, Y., {Inoue}, A.~K., \& {Sugahara}, Y. 2023, \mnras, 521, 2962,
  \dodoi{10.1093/mnras/stad743}

\bibitem[{{Gilda} {et~al.}(2021){Gilda}, {Lower}, \& {Narayanan}}]{gilda_2021}
{Gilda}, S., {Lower}, S., \& {Narayanan}, D. 2021, \apj, 916, 43,
  \dodoi{10.3847/1538-4357/ac0058}

\bibitem[{{Graziani} {et~al.}(2020){Graziani}, {Schneider}, {Ginolfi}, {Hunt},
  {Maio}, {Glatzle}, \& {Ciardi}}]{graziani_2020}
{Graziani}, L., {Schneider}, R., {Ginolfi}, M., {et~al.} 2020, \mnras, 494,
  1071, \dodoi{10.1093/mnras/staa796}

\bibitem[{{Greve} {et~al.}(2012){Greve}, {Vieira}, {Wei{\ss}}, {Aguirre},
  {Aird}, {Ashby}, {Benson}, {Bleem}, {Bradford}, {Brodwin}, {Carlstrom},
  {Chang}, {Chapman}, {Crawford}, {de Breuck}, {de Haan}, {Dobbs}, {Downes},
  {Fassnacht}, {Fazio}, {George}, {Gladders}, {Gonzalez}, {Halverson},
  {Hezaveh}, {High}, {Holder}, {Holzapfel}, {Hoover}, {Hrubes}, {Johnson},
  {Keisler}, {Knox}, {Lee}, {Leitch}, {Lueker}, {Luong-Van}, {Malkan},
  {Marrone}, {McIntyre}, {McMahon}, {Mehl}, {Menten}, {Meyer}, {Montroy},
  {Murphy}, {Natoli}, {Padin}, {Plagge}, {Pryke}, {Reichardt}, {Rest},
  {Rosenman}, {Ruel}, {Ruhl}, {Schaffer}, {Sharon}, {Shaw}, {Shirokoff},
  {Stalder}, {Stanford}, {Staniszewski}, {Stark}, {Story}, {Vanderlinde},
  {Walsh}, {Welikala}, \& {Williamson}}]{greve_2012}
{Greve}, T.~R., {Vieira}, J.~D., {Wei{\ss}}, A., {et~al.} 2012, \apj, 756, 101,
  \dodoi{10.1088/0004-637X/756/1/101}

\bibitem[{{Hahn} \& {Abel}(2011)}]{hahn_abel_2011_music}
{Hahn}, O., \& {Abel}, T. 2011, \mnras, 415, 2101,
  \dodoi{10.1111/j.1365-2966.2011.18820.x}

\bibitem[{{Harikane} {et~al.}(2020){Harikane}, {Ouchi}, {Inoue}, {Matsuoka},
  {Tamura}, {Bakx}, {Fujimoto}, {Moriwaki}, {Ono}, {Nagao}, {Tadaki}, {Kojima},
  {Shibuya}, {Egami}, {Ferrara}, {Gallerani}, {Hashimoto}, {Kohno}, {Matsuda},
  {Matsuo}, {Pallottini}, {Sugahara}, \& {Vallini}}]{harikane_2020}
{Harikane}, Y., {Ouchi}, M., {Inoue}, A.~K., {et~al.} 2020, \apj, 896, 93,
  \dodoi{10.3847/1538-4357/ab94bd}

\bibitem[{Harris {et~al.}(2020)Harris, Millman, van~der Walt, Gommers,
  Virtanen, Cournapeau, Wieser, Taylor, Berg, Smith, Kern, Picus, Hoyer, van
  Kerkwijk, Brett, Haldane, del R{\'{i}}o, Wiebe, Peterson,
  G{\'{e}}rard-Marchant, Sheppard, Reddy, Weckesser, Abbasi, Gohlke, \&
  Oliphant}]{harris2020array_numpy}
Harris, C.~R., Millman, K.~J., van~der Walt, S.~J., {et~al.} 2020, Nature, 585,
  357, \dodoi{10.1038/s41586-020-2649-2}

\bibitem[{{Hashimoto} {et~al.}(2019){Hashimoto}, {Inoue}, {Mawatari}, {Tamura},
  {Matsuo}, {Furusawa}, {Harikane}, {Shibuya}, {Knudsen}, {Kohno}, {Ono},
  {Zackrisson}, {Okamoto}, {Kashikawa}, {Oesch}, {Ouchi}, {Ota}, {Shimizu},
  {Taniguchi}, {Umehata}, \& {Watson}}]{hashimoto_2019}
{Hashimoto}, T., {Inoue}, A.~K., {Mawatari}, K., {et~al.} 2019, \pasj, 71, 71,
  \dodoi{10.1093/pasj/psz049}

\bibitem[{{Hashimoto} {et~al.}(2022){Hashimoto}, {Inoue}, {Sugahara},
  {Fudamoto}, {Fujimoto}, {Knudsen}, {Matsuo}, {Tamura}, {Yamanaka},
  {Harikane}, {Kuno}, {Ono}, \& {Salak}}]{hashimoto_2022}
{Hashimoto}, T., {Inoue}, A.~K., {Sugahara}, Y., {et~al.} 2022, arXiv e-prints,
  arXiv:2203.01345, \dodoi{10.48550/arXiv.2203.01345}

\bibitem[{{Haskell} {et~al.}(2023){Haskell}, {Smith}, {Cochrane}, {Hayward}, \&
  {Angl{\'e}s-Alc{\'a}zar}}]{haskell_2023}
{Haskell}, P., {Smith}, D.~J.~B., {Cochrane}, R.~K., {Hayward}, C.~C., \&
  {Angl{\'e}s-Alc{\'a}zar}, D. 2023, \mnras, 525, 1535,
  \dodoi{10.1093/mnras/stad2315}

\bibitem[{{Hildebrand}(1983)}]{hildebrand_1983}
{Hildebrand}, R.~H. 1983, \qjras, 24, 267

\bibitem[{{Hopkins}(2015)}]{hopkins_2015_gizmo}
{Hopkins}, P.~F. 2015, \mnras, 450, 53, \dodoi{10.1093/mnras/stv195}

\bibitem[{{Hopkins} {et~al.}(2014){Hopkins}, {Kere{\v{s}}}, {O{\~n}orbe},
  {Faucher-Gigu{\`e}re}, {Quataert}, {Murray}, \& {Bullock}}]{hopkins_FIRE_og}
{Hopkins}, P.~F., {Kere{\v{s}}}, D., {O{\~n}orbe}, J., {et~al.} 2014, \mnras,
  445, 581, \dodoi{10.1093/mnras/stu1738}

\bibitem[{Hunter(2007)}]{Hunter_2007}
Hunter, J.~D. 2007, Computing in Science \& Engineering, 9, 90,
  \dodoi{10.1109/MCSE.2007.55}

\bibitem[{{Inami} {et~al.}(2022){Inami}, {Algera}, {Schouws}, {Sommovigo},
  {Bouwens}, {Smit}, {Stefanon}, {Bowler}, {Endsley}, {Ferrara}, {Oesch},
  {Stark}, {Aravena}, {Barrufet}, {da Cunha}, {Dayal}, {De Looze}, {Fudamoto},
  {Gonzalez}, {Graziani}, {Hodge}, {Hygate}, {Nanayakkara}, {Pallottini},
  {Riechers}, {Schneider}, {Topping}, \& {van der Werf}}]{inami_2022}
{Inami}, H., {Algera}, H. S.~B., {Schouws}, S., {et~al.} 2022, \mnras, 515,
  3126, \dodoi{10.1093/mnras/stac1779}

\bibitem[{{Inoue} {et~al.}(2020){Inoue}, {Hashimoto}, {Chihara}, \&
  {Koike}}]{inoue_2020}
{Inoue}, A.~K., {Hashimoto}, T., {Chihara}, H., \& {Koike}, C. 2020, \mnras,
  495, 1577, \dodoi{10.1093/mnras/staa1203}

\bibitem[{{Iwamoto} {et~al.}(1999){Iwamoto}, {Brachwitz}, {Nomoto},
  {Kishimoto}, {Umeda}, {Hix}, \& {Thielemann}}]{iwamoto_1999_sne_yields}
{Iwamoto}, K., {Brachwitz}, F., {Nomoto}, K., {et~al.} 1999, \apjs, 125, 439,
  \dodoi{10.1086/313278}

\bibitem[{{Jin} {et~al.}(2022){Jin}, {Daddi}, {Magdis}, {Liu}, {Weaver}, {Tan},
  {Valentino}, {Gao}, {Schinnerer}, {Calabr{\`o}}, {Gu}, \& {Sese}}]{jin_2022}
{Jin}, S., {Daddi}, E., {Magdis}, G.~E., {et~al.} 2022, \aap, 665, A3,
  \dodoi{10.1051/0004-6361/202243341}

\bibitem[{{Johnson} {et~al.}(2021){Johnson}, {Leja}, {Conroy}, \&
  {Speagle}}]{johnson_2021}
{Johnson}, B.~D., {Leja}, J., {Conroy}, C., \& {Speagle}, J.~S. 2021, \apjs,
  254, 22, \dodoi{10.3847/1538-4365/abef67}

\bibitem[{{Jones} \& {Stanway}(2023)}]{jones_2023}
{Jones}, G.~T., \& {Stanway}, E.~R. 2023, \mnras, 525, 5720,
  \dodoi{10.1093/mnras/stad2683}

\bibitem[{{Jones} {et~al.}(2022){Jones}, {Stanway}, \& {Carnall}}]{jones_2022}
{Jones}, G.~T., {Stanway}, E.~R., \& {Carnall}, A.~C. 2022, \mnras, 514, 5706,
  \dodoi{10.1093/mnras/stac1667}

\bibitem[{{Juvela} {et~al.}(2013){Juvela}, {Montillaud}, {Ysard}, \&
  {Lunttila}}]{juvela_2013}
{Juvela}, M., {Montillaud}, J., {Ysard}, N., \& {Lunttila}, T. 2013, \aap, 556,
  A63, \dodoi{10.1051/0004-6361/201220910}

\bibitem[{{Kelly} {et~al.}(2012){Kelly}, {Shetty}, {Stutz}, {Kauffmann},
  {Goodman}, \& {Launhardt}}]{kelly_2012}
{Kelly}, B.~C., {Shetty}, R., {Stutz}, A.~M., {et~al.} 2012, \apj, 752, 55,
  \dodoi{10.1088/0004-637X/752/1/55}

\bibitem[{{Kirkpatrick} {et~al.}(2014){Kirkpatrick}, {Calzetti}, {Kennicutt},
  {Galametz}, {Gordon}, {Groves}, {Hunt}, {Dale}, {Hinz}, \&
  {Tabatabaei}}]{kirkpatrick_2014}
{Kirkpatrick}, A., {Calzetti}, D., {Kennicutt}, R., {et~al.} 2014, \apj, 789,
  130, \dodoi{10.1088/0004-637X/789/2/130}

\bibitem[{{Klaas} {et~al.}(2001){Klaas}, {Haas}, {M{\"u}ller}, {Chini},
  {Schulz}, {Coulson}, {Hippelein}, {Wilke}, {Albrecht}, \&
  {Lemke}}]{klaas_2001}
{Klaas}, U., {Haas}, M., {M{\"u}ller}, S.~A.~H., {et~al.} 2001, \aap, 379, 823,
  \dodoi{10.1051/0004-6361:20011377}

\bibitem[{{Knudsen} {et~al.}(2017){Knudsen}, {Watson}, {Frayer}, {Christensen},
  {Gallazzi}, {Micha{\l}owski}, {Richard}, \& {Zavala}}]{knudsen_2017}
{Knudsen}, K.~K., {Watson}, D., {Frayer}, D., {et~al.} 2017, \mnras, 466, 138,
  \dodoi{10.1093/mnras/stw3066}

\bibitem[{{Kov{\'a}cs} {et~al.}(2010){Kov{\'a}cs}, {Omont}, {Beelen},
  {Lonsdale}, {Polletta}, {Fiolet}, {Greve}, {Borys}, {Cox}, {De Breuck},
  {Dole}, {Dowell}, {Farrah}, {Lagache}, {Menten}, {Bell}, \&
  {Owen}}]{kovacs_2010}
{Kov{\'a}cs}, A., {Omont}, A., {Beelen}, A., {et~al.} 2010, \apj, 717, 29,
  \dodoi{10.1088/0004-637X/717/1/29}

\bibitem[{{Kroupa}(2002)}]{kroupa_initial_2002}
{Kroupa}, P. 2002, Science, 295, 82, \dodoi{10.1126/science.1067524}

\bibitem[{{Krumholz} \& {Gnedin}(2011)}]{krumholz_2011_h2}
{Krumholz}, M.~R., \& {Gnedin}, N.~Y. 2011, \apj, 729, 36,
  \dodoi{10.1088/0004-637X/729/1/36}

\bibitem[{{Labb{\'e}} {et~al.}(2023){Labb{\'e}}, {van Dokkum}, {Nelson},
  {Bezanson}, {Suess}, {Leja}, {Brammer}, {Whitaker}, {Mathews}, {Stefanon}, \&
  {Wang}}]{labbe_2023}
{Labb{\'e}}, I., {van Dokkum}, P., {Nelson}, E., {et~al.} 2023, \nat, 616, 266,
  \dodoi{10.1038/s41586-023-05786-2}

\bibitem[{{Le F{\`e}vre} {et~al.}(2020){Le F{\`e}vre}, {B{\'e}thermin},
  {Faisst}, {Jones}, {Capak}, {Cassata}, {Silverman}, {Schaerer}, {Yan},
  {Amorin}, {Bardelli}, {Boquien}, {Cimatti}, {Dessauges-Zavadsky},
  {Giavalisco}, {Hathi}, {Fudamoto}, {Fujimoto}, {Ginolfi}, {Gruppioni},
  {Hemmati}, {Ibar}, {Koekemoer}, {Khusanova}, {Lagache}, {Lemaux}, {Loiacono},
  {Maiolino}, {Mancini}, {Narayanan}, {Morselli}, {M{\'e}ndez-Hern{\`a}ndez},
  {Oesch}, {Pozzi}, {Romano}, {Riechers}, {Scoville}, {Talia}, {Tasca},
  {Thomas}, {Toft}, {Vallini}, {Vergani}, {Walter}, {Zamorani}, \&
  {Zucca}}]{le_fevre_2020_alpine}
{Le F{\`e}vre}, O., {B{\'e}thermin}, M., {Faisst}, A., {et~al.} 2020, \aap,
  643, A1, \dodoi{10.1051/0004-6361/201936965}

\bibitem[{{Lee} {et~al.}(2015){Lee}, {Leroy}, {Schnee}, {Wong}, {Bolatto},
  {Indebetouw}, \& {Rubio}}]{lee_2015}
{Lee}, C., {Leroy}, A.~K., {Schnee}, S., {et~al.} 2015, \mnras, 450, 2708,
  \dodoi{10.1093/mnras/stv863}

\bibitem[{{Lewis} {et~al.}(2022){Lewis}, {Ocvirk}, {Dubois}, {Aubert},
  {Chardin}, {Gillet}, \& {Th{\'e}lie}}]{lewis_2022_dustier}
{Lewis}, J. S.~W., {Ocvirk}, P., {Dubois}, Y., {et~al.} 2022, arXiv e-prints,
  arXiv:2204.03949.
\newblock \doarXiv{2204.03949}

\bibitem[{{Li} \& {Draine}(2001)}]{li_draine_01_2}
{Li}, A., \& {Draine}, B.~T. 2001, \apj, 554, 778, \dodoi{10.1086/323147}

\bibitem[{{Li} {et~al.}(2019){Li}, {Narayanan}, \& {Dav{\'e}}}]{li_2019_dust}
{Li}, Q., {Narayanan}, D., \& {Dav{\'e}}, R. 2019, \mnras, 490, 1425,
  \dodoi{10.1093/mnras/stz2684}

\bibitem[{{Li} {et~al.}(2021){Li}, {Narayanan}, {Torrey}, {Dav{\'e}}, \&
  {Vogelsberger}}]{li_2020_mw_ext_curve}
{Li}, Q., {Narayanan}, D., {Torrey}, P., {Dav{\'e}}, R., \& {Vogelsberger}, M.
  2021, \mnras, 507, 548, \dodoi{10.1093/mnras/stab2196}

\bibitem[{{Liang} {et~al.}(2019){Liang}, {Feldmann}, {Kere{\v{s}}}, {Scoville},
  {Hayward}, {Faucher-Gigu{\`e}re}, {Schreiber}, {Ma}, {Hopkins}, \&
  {Quataert}}]{liang_lichen_2019}
{Liang}, L., {Feldmann}, R., {Kere{\v{s}}}, D., {et~al.} 2019, \mnras, 489,
  1397, \dodoi{10.1093/mnras/stz2134}

\bibitem[{{Lovell} {et~al.}(2022){Lovell}, {Geach}, {Dav{\'e}}, {Narayanan},
  {Coppin}, {Li}, {Franco}, \& {Privon}}]{lovell_2022_orientation}
{Lovell}, C.~C., {Geach}, J.~E., {Dav{\'e}}, R., {et~al.} 2022, \mnras, 515,
  3644, \dodoi{10.1093/mnras/stac2008}

\bibitem[{{Lower} {et~al.}(2022){Lower}, {Narayanan}, {Li}, \&
  {Dav{\'e}}}]{lower_2022_cosmic_sands}
{Lower}, S., {Narayanan}, D., {Li}, Q., \& {Dav{\'e}}, R. 2022, arXiv e-prints,
  arXiv:2212.02636.
\newblock \doarXiv{2212.02636}

\bibitem[{{Lucy}(1999)}]{lucy_rt}
{Lucy}, L.~B. 1999, \aap, 344, 282

\bibitem[{{Marinacci} {et~al.}(2019){Marinacci}, {Sales}, {Vogelsberger},
  {Torrey}, \& {Springel}}]{marinacci_2019_smuggle}
{Marinacci}, F., {Sales}, L.~V., {Vogelsberger}, M., {Torrey}, P., \&
  {Springel}, V. 2019, \mnras, 489, 4233,
  \dodoi{10.1093/mnras/stz239110.48550/arXiv.1905.08806}

\bibitem[{{McKinnon} {et~al.}(2016){McKinnon}, {Torrey}, \&
  {Vogelsberger}}]{mckinnon_2016}
{McKinnon}, R., {Torrey}, P., \& {Vogelsberger}, M. 2016, \mnras, 457, 3775,
  \dodoi{10.1093/mnras/stw253}

\bibitem[{{Muratov} {et~al.}(2015){Muratov}, {Kere{\v{s}}},
  {Faucher-Gigu{\`e}re}, {Hopkins}, {Quataert}, \& {Murray}}]{muratov_2015}
{Muratov}, A.~L., {Kere{\v{s}}}, D., {Faucher-Gigu{\`e}re}, C.-A., {et~al.}
  2015, \mnras, 454, 2691, \dodoi{10.1093/mnras/stv2126}

\bibitem[{{Mushtaq} \& {Puttasiddappa}(2022)}]{mushtaq_2022}
{Mushtaq}, M., \& {Puttasiddappa}, P.~H. 2022, arXiv e-prints,
  arXiv:2208.08658, \dodoi{10.48550/arXiv.2208.08658}

\bibitem[{{Nagaraj} {et~al.}(2022){Nagaraj}, {Forbes}, {Leja},
  {Foreman-Mackey}, \& {Hayward}}]{nagaraj_2022}
{Nagaraj}, G., {Forbes}, J.~C., {Leja}, J., {Foreman-Mackey}, D., \& {Hayward},
  C.~C. 2022, \apj, 932, 54, \dodoi{10.3847/1538-4357/ac6c80}

\bibitem[{{Narayanan} {et~al.}(2020){Narayanan}, {Turk}, {Robitaille}, {Kelly},
  {Connor McClellan}, {Sharma}, {Garg}, {Abruzzo}, {Choi}, {Conroy}, {Johnson},
  {Kimock}, {Li}, {Lovell}, {Lower}, {Privon}, {Roberts}, {Sethuram}, {Snyder},
  {Thompson}, \& {Wise}}]{powderday}
{Narayanan}, D., {Turk}, M.~J., {Robitaille}, T., {et~al.} 2020, arXiv
  e-prints, arXiv:2006.10757.
\newblock \doarXiv{2006.10757}

\bibitem[{{Narayanan} {et~al.}(2023){Narayanan}, {Smith}, {Hensley}, {Li},
  {Hu}, {Sandstrom}, {Torrey}, {Vogelsberger}, {Marinacci}, \&
  {Sales}}]{narayanan_2023}
{Narayanan}, D., {Smith}, J.~D., {Hensley}, B., {et~al.} 2023, arXiv e-prints,
  arXiv:2301.07136, \dodoi{10.48550/arXiv.2301.07136}

\bibitem[{{Nomoto} {et~al.}(2006){Nomoto}, {Tominaga}, {Umeda}, {Kobayashi}, \&
  {Maeda}}]{nomoto_simba_sne_yields}
{Nomoto}, K., {Tominaga}, N., {Umeda}, H., {Kobayashi}, C., \& {Maeda}, K.
  2006, \nphysa, 777, 424, \dodoi{10.1016/j.nuclphysa.2006.05.008}

\bibitem[{{Oppenheimer} \& {Dav{\'e}}(2006)}]{Oppenheimer_2006_agb_yields}
{Oppenheimer}, B.~D., \& {Dav{\'e}}, R. 2006, \mnras, 373, 1265,
  \dodoi{10.1111/j.1365-2966.2006.10989.x}

\bibitem[{pandas~development team(2020)}]{reback2020pandas}
pandas~development team, T. 2020, pandas-dev/pandas: Pandas, latest,  Zenodo,
  \dodoi{10.5281/zenodo.3509134}

\bibitem[{{Pavesi} {et~al.}(2016){Pavesi}, {Riechers}, {Capak}, {Carilli},
  {Sharon}, {Stacey}, {Karim}, {Scoville}, \&
  {Smol{\v{c}}i{\'c}}}]{pavesi_2016}
{Pavesi}, R., {Riechers}, D.~A., {Capak}, P.~L., {et~al.} 2016, \apj, 832, 151,
  \dodoi{10.3847/0004-637X/832/2/151}

\bibitem[{{Rahmati} {et~al.}(2013){Rahmati}, {Pawlik}, {Rai{\v{c}}evi{\'c}}, \&
  {Schaye}}]{rahmati_2013_self_shield}
{Rahmati}, A., {Pawlik}, A.~H., {Rai{\v{c}}evi{\'c}}, M., \& {Schaye}, J. 2013,
  \mnras, 430, 2427, \dodoi{10.1093/mnras/stt066}

\bibitem[{{Rangwala} {et~al.}(2011){Rangwala}, {Maloney}, {Glenn}, {Wilson},
  {Rykala}, {Isaak}, {Baes}, {Bendo}, {Boselli}, {Bradford}, {Clements},
  {Cooray}, {Fulton}, {Imhof}, {Kamenetzky}, {Madden}, {Mentuch}, {Sacchi},
  {Sauvage}, {Schirm}, {Smith}, {Spinoglio}, \& {Wolfire}}]{rangwala_2011}
{Rangwala}, N., {Maloney}, P.~R., {Glenn}, J., {et~al.} 2011, \apj, 743, 94,
  \dodoi{10.1088/0004-637X/743/1/94}

\bibitem[{{Reuter} {et~al.}(2020){Reuter}, {Vieira}, {Spilker}, {Weiss},
  {Aravena}, {Archipley}, {B{\'e}thermin}, {Chapman}, {De Breuck}, {Dong},
  {Everett}, {Fu}, {Greve}, {Hayward}, {Hill}, {Hezaveh}, {Jarugula}, {Litke},
  {Malkan}, {Marrone}, {Narayanan}, {Phadke}, {Stark}, \&
  {Strandet}}]{reuter_2020}
{Reuter}, C., {Vieira}, J.~D., {Spilker}, J.~S., {et~al.} 2020, \apj, 902, 78,
  \dodoi{10.3847/1538-4357/abb599}

\bibitem[{{Reuter} {et~al.}(2023){Reuter}, {Spilker}, {Vieira}, {Marrone},
  {Weiss}, {Aravena}, {Archipley}, {Chapman}, {Gonzalez}, {Greve}, {Hayward},
  {Hill}, {Jarugula}, {Kim}, {Malkan}, {Phadke}, {Stark}, {Sulzenauer}, \&
  {Vizgan}}]{reuter_2023}
{Reuter}, C., {Spilker}, J.~S., {Vieira}, J.~D., {et~al.} 2023, \apj, 948, 44,
  \dodoi{10.3847/1538-4357/acaf51}

\bibitem[{{Robitaille}(2011)}]{robitaille_2011_hyperion}
{Robitaille}, T.~P. 2011, \aap, 536, A79, \dodoi{10.1051/0004-6361/201117150}

\bibitem[{{Robitaille} {et~al.}(2012){Robitaille}, {Churchwell}, {Benjamin},
  {Whitney}, {Wood}, {Babler}, \& {Meade}}]{robitaille_pahs}
{Robitaille}, T.~P., {Churchwell}, E., {Benjamin}, R.~A., {et~al.} 2012, \aap,
  545, A39, \dodoi{10.1051/0004-6361/201219073}

\bibitem[{{Salvatier} {et~al.}(2016){Salvatier}, {Wiecki{\^a}}, \&
  {Fonnesbeck}}]{salvatier_2016}
{Salvatier}, J., {Wiecki{\^a}}, T.~V., \& {Fonnesbeck}, C. 2016, {PyMC3: Python
  probabilistic programming framework}, Astrophysics Source Code Library,
  record ascl:1610.016.
\newblock \doeprint{1610.016}

\bibitem[{{Scoville} {et~al.}(2017){Scoville}, {Lee}, {Vanden Bout},
  {Diaz-Santos}, {Sanders}, {Darvish}, {Bongiorno}, {Casey}, {Murchikova},
  {Koda}, {Capak}, {Vlahakis}, {Ilbert}, {Sheth}, {Morokuma-Matsui}, {Ivison},
  {Aussel}, {Laigle}, {McCracken}, {Armus}, {Pope}, {Toft}, \&
  {Masters}}]{scoville_2017}
{Scoville}, N., {Lee}, N., {Vanden Bout}, P., {et~al.} 2017, \apj, 837, 150,
  \dodoi{10.3847/1538-4357/aa61a0}

\bibitem[{{Shen} {et~al.}(2022){Shen}, {Vogelsberger}, {Nelson}, {Tacchella},
  {Hernquist}, {Springel}, {Marinacci}, \& {Torrey}}]{shen_2022_tng}
{Shen}, X., {Vogelsberger}, M., {Nelson}, D., {et~al.} 2022, \mnras, 510, 5560,
  \dodoi{10.1093/mnras/stab3794}

\bibitem[{{Shetty} {et~al.}(2009{\natexlab{a}}){Shetty}, {Kauffmann}, {Schnee},
  \& {Goodman}}]{shetty_2009a}
{Shetty}, R., {Kauffmann}, J., {Schnee}, S., \& {Goodman}, A.~A.
  2009{\natexlab{a}}, \apj, 696, 676, \dodoi{10.1088/0004-637X/696/1/676}

\bibitem[{{Shetty} {et~al.}(2009{\natexlab{b}}){Shetty}, {Kauffmann}, {Schnee},
  {Goodman}, \& {Ercolano}}]{shetty_2009b}
{Shetty}, R., {Kauffmann}, J., {Schnee}, S., {Goodman}, A.~A., \& {Ercolano},
  B. 2009{\natexlab{b}}, \apj, 696, 2234, \dodoi{10.1088/0004-637X/696/2/2234}

\bibitem[{{Smith} {et~al.}(2017){Smith}, {Bryan}, {Glover}, {Goldbaum}, {Turk},
  {Regan}, {Wise}, {Schive}, {Abel}, {Emerick}, {O'Shea}, {Anninos}, {Hummels},
  \& {Khochfar}}]{smith_2017_grackle}
{Smith}, B.~D., {Bryan}, G.~L., {Glover}, S. C.~O., {et~al.} 2017, \mnras, 466,
  2217, \dodoi{10.1093/mnras/stw3291}

\bibitem[{{Sommovigo} {et~al.}(2021){Sommovigo}, {Ferrara}, {Carniani},
  {Zanella}, {Pallottini}, {Gallerani}, \& {Vallini}}]{sommovigo_2021}
{Sommovigo}, L., {Ferrara}, A., {Carniani}, S., {et~al.} 2021, \mnras, 503,
  4878, \dodoi{10.1093/mnras/stab720}

\bibitem[{{Sommovigo} {et~al.}(2022{\natexlab{a}}){Sommovigo}, {Ferrara},
  {Pallottini}, {Dayal}, {Bouwens}, {Smit}, {da Cunha}, {De Looze}, {Bowler},
  {Hodge}, {Inami}, {Oesch}, {Endsley}, {Gonzalez}, {Schouws}, {Stark},
  {Stefanon}, {Aravena}, {Graziani}, {Riechers}, {Schneider}, {van der Werf},
  {Algera}, {Barrufet}, {Fudamoto}, {Hygate}, {Labb{\'e}}, {Li}, {Nanayakkara},
  \& {Topping}}]{sommovigo_2022_rebels_temps}
{Sommovigo}, L., {Ferrara}, A., {Pallottini}, A., {et~al.} 2022{\natexlab{a}},
  \mnras, 513, 3122, \dodoi{10.1093/mnras/stac302}

\bibitem[{{Sommovigo} {et~al.}(2022{\natexlab{b}}){Sommovigo}, {Ferrara},
  {Carniani}, {Pallottini}, {Dayal}, {Pizzati}, {Ginolfi}, {Markov}, \&
  {Faisst}}]{sommovigo_2022_rebels_alpine_dust}
{Sommovigo}, L., {Ferrara}, A., {Carniani}, S., {et~al.} 2022{\natexlab{b}},
  \mnras, 517, 5930, \dodoi{10.1093/mnras/stac2997}

\bibitem[{{Spilker} {et~al.}(2016){Spilker}, {Marrone}, {Aravena},
  {B{\'e}thermin}, {Bothwell}, {Carlstrom}, {Chapman}, {Crawford}, {de Breuck},
  {Fassnacht}, {Gonzalez}, {Greve}, {Hezaveh}, {Litke}, {Ma}, {Malkan},
  {Rotermund}, {Strandet}, {Vieira}, {Weiss}, \& {Welikala}}]{spilker_2016}
{Spilker}, J.~S., {Marrone}, D.~P., {Aravena}, M., {et~al.} 2016, \apj, 826,
  112, \dodoi{10.3847/0004-637X/826/2/112}

\bibitem[{{Spilker} {et~al.}(2022){Spilker}, {Hayward}, {Marrone}, {Aravena},
  {B{\'e}thermin}, {Burgoyne}, {Chapman}, {Greve}, {Gururajan}, {Hezaveh},
  {Hill}, {Litke}, {Lovell}, {Malkan}, {Murphy}, {Narayanan}, {Phadke},
  {Reuter}, {Stark}, {Sulzenauer}, {Vieira}, {Vizgan}, \&
  {Wei{\ss}}}]{spilker_2022}
{Spilker}, J.~S., {Hayward}, C.~C., {Marrone}, D.~P., {et~al.} 2022, \apjl,
  929, L3, \dodoi{10.3847/2041-8213/ac61e6}

\bibitem[{{Strandet} {et~al.}(2016){Strandet}, {Weiss}, {Vieira}, {de Breuck},
  {Aguirre}, {Aravena}, {Ashby}, {B{\'e}thermin}, {Bradford}, {Carlstrom},
  {Chapman}, {Crawford}, {Everett}, {Fassnacht}, {Furstenau}, {Gonzalez},
  {Greve}, {Gullberg}, {Hezaveh}, {Kamenetzky}, {Litke}, {Ma}, {Malkan},
  {Marrone}, {Menten}, {Murphy}, {Nadolski}, {Rotermund}, {Spilker}, {Stark},
  \& {Welikala}}]{strandet_2016}
{Strandet}, M.~L., {Weiss}, A., {Vieira}, J.~D., {et~al.} 2016, \apj, 822, 80,
  \dodoi{10.3847/0004-637X/822/2/80}

\bibitem[{{Sun} {et~al.}(2022){Sun}, {Egami}, {Fujimoto}, {Rawle}, {Bauer},
  {Kohno}, {Smail}, {P{\'e}rez-Gonz{\'a}lez}, {Ao}, {Chapman}, {Combes},
  {Dessauges-Zavadsky}, {Espada}, {Gonz{\'a}lez-L{\'o}pez}, {Koekemoer},
  {Kokorev}, {Lee}, {Morokuma-Matsui}, {Mu{\~n}oz Arancibia}, {Oguri},
  {Pell{\'o}}, {Ueda}, {Uematsu}, {Valentino}, {Van der Werf}, {Walth},
  {Zemcov}, \& {Zitrin}}]{sun_2022}
{Sun}, F., {Egami}, E., {Fujimoto}, S., {et~al.} 2022, \apj, 932, 77,
  \dodoi{10.3847/1538-4357/ac6e3f}

\bibitem[{{Swinbank} {et~al.}(2010){Swinbank}, {Smail}, {Longmore}, {Harris},
  {Baker}, {De Breuck}, {Richard}, {Edge}, {Ivison}, {Blundell}, {Coppin},
  {Cox}, {Gurwell}, {Hainline}, {Krips}, {Lundgren}, {Neri}, {Siana},
  {Siringo}, {Stark}, {Wilner}, \& {Younger}}]{swinbank_2010}
{Swinbank}, A.~M., {Smail}, I., {Longmore}, S., {et~al.} 2010, \nat, 464, 733,
  \dodoi{10.1038/nature08880}

\bibitem[{{Swinbank} {et~al.}(2014){Swinbank}, {Simpson}, {Smail}, {Harrison},
  {Hodge}, {Karim}, {Walter}, {Alexander}, {Brandt}, {de Breuck}, {da Cunha},
  {Chapman}, {Coppin}, {Danielson}, {Dannerbauer}, {Decarli}, {Greve},
  {Ivison}, {Knudsen}, {Lagos}, {Schinnerer}, {Thomson}, {Wardlow}, {Wei{\ss}},
  \& {van der Werf}}]{swinbank_2014}
{Swinbank}, A.~M., {Simpson}, J.~M., {Smail}, I., {et~al.} 2014, \mnras, 438,
  1267, \dodoi{10.1093/mnras/stt2273}

\bibitem[{{Tamura} {et~al.}(2019){Tamura}, {Mawatari}, {Hashimoto}, {Inoue},
  {Zackrisson}, {Christensen}, {Binggeli}, {Matsuda}, {Matsuo}, {Takeuchi},
  {Asano}, {Sunaga}, {Shimizu}, {Okamoto}, {Yoshida}, {Lee}, {Shibuya},
  {Taniguchi}, {Umehata}, {Hatsukade}, {Kohno}, \& {Ota}}]{tamura_2019}
{Tamura}, Y., {Mawatari}, K., {Hashimoto}, T., {et~al.} 2019, \apj, 874, 27,
  \dodoi{10.3847/1538-4357/ab0374}

\bibitem[{{Thompson}(2014)}]{caesar}
{Thompson}, R. 2014, {pyGadgetReader: GADGET snapshot reader for python}.
\newblock \doeprint{1411.001}

\bibitem[{{Turk} {et~al.}(2011){Turk}, {Smith}, {Oishi}, {Skory}, {Skillman},
  {Abel}, \& {Norman}}]{turk_2011_yt}
{Turk}, M.~J., {Smith}, B.~D., {Oishi}, J.~S., {et~al.} 2011, \apjs, 192, 9,
  \dodoi{10.1088/0067-0049/192/1/9}

\bibitem[{{Utomo} {et~al.}(2019){Utomo}, {Chiang}, {Leroy}, {Sandstrom}, \&
  {Chastenet}}]{utomo_2019}
{Utomo}, D., {Chiang}, I.-D., {Leroy}, A.~K., {Sandstrom}, K.~M., \&
  {Chastenet}, J. 2019, \apj, 874, 141, \dodoi{10.3847/1538-4357/ab05d3}

\bibitem[{{Vijayan} {et~al.}(2019){Vijayan}, {Clay}, {Thomas}, {Yates},
  {Wilkins}, \& {Henriques}}]{vijayan_2019}
{Vijayan}, A.~P., {Clay}, S.~J., {Thomas}, P.~A., {et~al.} 2019, \mnras, 489,
  4072, \dodoi{10.1093/mnras/stz1948}

\bibitem[{{Vijayan} {et~al.}(2022){Vijayan}, {Wilkins}, {Lovell}, {Thomas},
  {Camps}, {Baes}, {Trayford}, {Kuusisto}, \& {Roper}}]{vijayan_2022}
{Vijayan}, A.~P., {Wilkins}, S.~M., {Lovell}, C.~C., {et~al.} 2022, \mnras,
  511, 4999, \dodoi{10.1093/mnras/stac338}

\bibitem[{{Vizgan} {et~al.}(2022){Vizgan}, {Greve}, {Olsen}, {Zanella},
  {Narayanan}, {Dav{\`e}}, {Magdis}, {Popping}, {Valentino}, \&
  {Heintz}}]{vizgan_2022}
{Vizgan}, D., {Greve}, T.~R., {Olsen}, K.~P., {et~al.} 2022, \apj, 929, 92,
  \dodoi{10.3847/1538-4357/ac5cba}

\bibitem[{{Vogelsberger} {et~al.}(2020){Vogelsberger}, {Nelson}, {Pillepich},
  {Shen}, {Marinacci}, {Springel}, {Pakmor}, {Tacchella}, {Weinberger},
  {Torrey}, \& {Hernquist}}]{vogelsberger_2020}
{Vogelsberger}, M., {Nelson}, D., {Pillepich}, A., {et~al.} 2020, \mnras, 492,
  5167, \dodoi{10.1093/mnras/staa137}

\bibitem[{{Watson} {et~al.}(2015){Watson}, {Christensen}, {Knudsen}, {Richard},
  {Gallazzi}, \& {Micha{\l}owski}}]{watson_2015}
{Watson}, D., {Christensen}, L., {Knudsen}, K.~K., {et~al.} 2015, \nat, 519,
  327, \dodoi{10.1038/nature14164}

\bibitem[{{Weingartner} \& {Draine}(2001)}]{weingartner_draine_2001}
{Weingartner}, J.~C., \& {Draine}, B.~T. 2001, \apj, 548, 296,
  \dodoi{10.1086/318651}

\bibitem[{{W}es {M}c{K}inney(2010)}]{mckinney-proc-scipy-2010}
{W}es {M}c{K}inney. 2010, in {P}roceedings of the 9th {P}ython in {S}cience
  {C}onference, ed. {S}t\'efan van~der {W}alt \& {J}arrod {M}illman, 56 -- 61,
  \dodoi{10.25080/Majora-92bf1922-00a}

\bibitem[{{Witstok} {et~al.}(2023){Witstok}, {Jones}, {Maiolino}, {Smit}, \&
  {Schneider}}]{witstok_2023}
{Witstok}, J., {Jones}, G.~C., {Maiolino}, R., {Smit}, R., \& {Schneider}, R.
  2023, arXiv e-prints, arXiv:2305.09714, \dodoi{10.48550/arXiv.2305.09714}

\bibitem[{{Yajima} {et~al.}(2014){Yajima}, {Nagamine}, {Thompson}, \&
  {Choi}}]{yajima_2014}
{Yajima}, H., {Nagamine}, K., {Thompson}, R., \& {Choi}, J.-H. 2014, \mnras,
  439, 3073, \dodoi{10.1093/mnras/stu169}

\bibitem[{{Yang} \& {Phillips}(2007)}]{yang_2007}
{Yang}, M., \& {Phillips}, T. 2007, \apj, 662, 284, \dodoi{10.1086/514810}

\bibitem[{{Yoon} {et~al.}(2022){Yoon}, {Carilli}, {Fujimoto}, {Castellano},
  {Merlin}, {Santini}, {Yun}, {Murphy}, {Jung}, {Casey}, {Finkelstein},
  {Papovich}, {Fontana}, {Treu}, \& {Letai}}]{yoon_2022}
{Yoon}, I., {Carilli}, C.~L., {Fujimoto}, S., {et~al.} 2022, arXiv e-prints,
  arXiv:2210.08413, \dodoi{10.48550/arXiv.2210.08413}

\end{thebibliography}
\bibliographystyle{aasjournal}

\end{document}